\documentclass[manuscript]{aastex}
\usepackage{rotating}

\shorttitle{Baldwin effect and eigenvector 1 in AGNs}
\shortauthors{L. \v C. Popovi\'c \& J. Kova\v cevi\'c}

\begin{document}


\title{Optical emission line properties of a sample of the broad-line AGNs: the Baldwin effect and eigenvector 1}


\author{Luka \v C. Popovi\' c\altaffilmark{1,2} and Jelena
Kova\v cevi\' c\altaffilmark{1,2}}

\affil{\altaffilmark{1}Group for Astrophysical Spectroscopy, Astronomical  Observatory,  Volgina  7, 11060  Belgrade, Serbia}
\affil{\altaffilmark{2}Isaac Newton Institute of Chile, Yugoslavia branch}
\email{lpopovic@aob.bg.ac.rs,jkovacevic@aob.bg.ac.rs}


\begin{abstract}
We divide a sample of 302 type-1 AGNs into two subsamples based on the  narrow line [\ion{O}{3}]/H$\beta_{NLR}$ ratio, since we expect that there will be a stronger
starburst (\ion{H}{2} region) contribution to the narrow line emission for  $R$=log([\ion{O}{3}]/H$\beta_{NLR})<0.5$. For the two samples we { find significant differences in correlations between} spectral properties of objects with $R<0.5$ and $R>0.5$.
 {We find similar differences when we divided the sample based on the FWHM ratios of [\ion{O}{3}] and broad H$\beta$ lines
($R_1$=log(FWHM[\ion{O}{3}]/FWHM H$\beta_{\rm broad})^>_<-0.8$), i.e. similar correlations between $R>0.5$ and $R_1<-0.8$ subsamples from one side and $R<0.5$ and $R_1>-0.8$ subsamples from the other side.} The most interesting difference is in the correlation between the  broad H$\beta$ FWHM and luminosity in the $R<0.5$
($R_1>-0.8$) sample that indicates a connection between the BLR kinematics and photoionization source.   We discuss possible effects which can cause such differences in spectral properties of two subsamples.
\end{abstract}



\section{Introduction}

Spectral properties of Active Galactic Nuclei (AGNs)  depend on  physical conditions and geometry of emission regions which are quite likely to be changing during AGN evolution { \citep[see e.g.][]{lt06,WZ07}}. To understand the complex nature of AGN, many authors investigated the correlations between different spectral properties, trying to find and explain their physical background \citep[see e.g.][and references therein]{bg92,w99,cr02,sh03,y04,g04,w06,wa09,lu09,kov10}.
\citet{bg92} (BG92)  found a number of AGN spectral properties in a sample of AGNs, as e.g. that  as the optical to X-ray slope and equivalent width (EW) of \ion{Fe}{2} increase, the EW of [\ion{O}{3}] decreases, as well as  FWHM\footnote{Full Width at Half-Maximum Intensity of the H$\beta$ broad component} of H$\beta$. These correlations  are part of the eigenvector 1 (EV1) in BG92 principal component analysis. The underlying physics driving EV1 is not yet completely understood.

On the other hand, the Baldwin effect  \citep{ba77,ba78} is present in a number of lines, i.e.  their line EW decreases with increasing continuum luminosity. The majority of the UV and optical lines show a Baldwin effect \citep{d02}, but for some lines there is no trend  with the continuum luminosity, or even an inverse Baldwin effect. The inverse Baldwin effect is found for the \ion{Fe}{2} optical lines \citep{nt07,kov10} and for H$\beta$ \citep{cr02,nt07}. \citet{kov10} found an inverse Baldwin effect for the broad H$\beta$ component (in FWHM H$\beta_{\rm broad}$ $<$3000 km s$^{-1}$ subsample), but on the other hand, the narrow H$\beta$ component shows a normal Baldwin effect as do the [\ion{O}{3}] lines. The origin of the Baldwin effect is an open question still under debate \citep[for review see][and reference therein]{gr01}. Some indications of connection between the BG92 EV1 correlation with the Baldwin effect are discussed in the literature \citep[see e.g.][]{bl04,kov10,zh11}.

There are some indications that the EV1 may be connected with AGN evolution \citep[see][]{ma03,g04,w06}, while evolution of AGNs is probably related with starburst regions \citep{lt06,m09,sa10}. Namely, it is possible that AGNs in an earlier phase of their evolution are composed of starburst (star-forming) regions and the central engine (AGN), and, during evolution, the starburst contribution becomes weaker and/or negligible   \citep{WW06,WW08,m09}.

Authors of recent papers \citep[see e.g.][]{y04,g04,lu09} used a large sample of SDSS (Sloan Digital Sky Survey) objects and applied Principal Component (PC) analysis on some spectral parameters. They noticed that PC projections depend on the sample properties; i.e., different divisions of sample by luminosity, [\ion{O}{3}] strength, FWHM H$\beta$ etc. lead to different correlations between  spectral parameters, sometimes even opposite correlations for different subsamples. Some authors  found that the significance of the Baldwin effect depends on the FWHM H$\beta$ range of the sample \citep[see e.g.][]{su09,za10,kov10}.

In this paper we investigate a sample of the broad-line AGNs. Using the ratio of the [OIII] and narrow H$\beta$ line we divide the sample in two groups and
consider the connection between the  BG92 EV1 and Baldwin effect into these groups separately.

The paper is organized as following: in \S2 we shortly describe the sample (subsamples) and methods of analysis, in \S 3 we present the obtained correlations and discuss the results, and in \S4 we outline our conclusions.

\section{The sample  and analysis}

The sample of 302 AGNs used in this paper is the same as { the one} presented in \citet{kov10}.  { The sample has been chosen from  the SDSS Release 7. It contains the broad-line AGNs within approximately uniform redshift range z=0--0.7, mainly within the luminosity range of 44$<$log($\lambda L_{5100}$)$<$46.}  Other details of the sample selection can be found in \citet{kov10}.

As is described in \citet{kov10}, we fitted the \ion{Fe}{2} lines with a { new template which enables precise estimation of  the \ion{Fe}{2} emission within the 4400-5500 \AA \ } wavelength range  (see Fig. \ref{f1}).  The Balmer lines were fit with three components: a narrow, an intermediate and a very broad component (H$\beta$ NLR, ILR and VBLR, respectively).
 The H$\beta$  broad component is taken as the sum of the H$\beta$ ILR and H$\beta$ VBLR components (see Fig. \ref{f09}). The FWHM and  FWM10\% (full width at 10\% of maximum) of the broad H$\beta$ line are measured as it shown in  Fig. \ref{f09}.

Both components of the [\ion{O}{3}] $\lambda\lambda$4959, 5007 \AA \ doublet originate from the same lower energy level and both have the negligible
optical depth since the transitions are strongly forbidden. We thus assumed that the [\ion{O}{3}] $\lambda$4959 \AA \ and [\ion{O}{3}] $\lambda$5007 \AA \ lines have the same emission-line profile; i.e. we fit each line of the doublet by one Gaussian (or by two, in the case of a significant asymmetry) assuming that the [\ion{O}{3}] $\lambda$4959 \AA \ and [\ion{O}{3}] $\lambda$5007 \AA \ lines have the same widths and shifts. We also fixed their intensity ratio at 2.99 \citep{d07}.

To estimate the NLR contribution to the total H$\beta$ flux (i.e., emission from the same region emitting the
[OIII] lines) we introduce following assumptions: (i) only one narrow H$\beta$ component is present in the total H$\beta$ emission and (ii)  this component  has the same kinematical parameters as the [\ion{O}{3}] lines, i.e. the same  width and shift as  the [\ion{O}{3}] lines. Very often a blue asymmetry in the [\ion{O}{3}] lines  was present and in such case  each of the [\ion{O}{3}] lines was fitted with two Gaussian components, but as a rule only one (central, non-shifted)  has the same kinematics as the narrow H$\beta$ line.

To plot a ''BPT diagram'' \citep{bpt} shown in Fig. \ref{f01}, we also estimated the narrow H$\alpha$ component, connecting the width and shift of the narrow component with ones of the [NII] lines.
As it can be seen from Fig.
\ref{f01},  a number of AGNs from the sample show a significant starburst contribution. It is interesting that these AGNs are mostly with FWHM H$\beta_{\rm broad}$ $<$ 3000 km s$^{-1}$ (open circles), while in the AGN part of the BPT diagram both  fractions (with FWHM H$\beta_{\rm broad}$ $<$ 3000 km s$^{-1}$ and FWHM H$\beta_{\rm broad}$ $>$ 3000 km s$^{-1}$ -- full circles) are present.  Note here that  \citet{sa10} also found more intense circumnuclear star-forming activity in Narrow Line Seyfert 1s (NLS1) than in the broad-line AGNs.

 { Since we have  complete measurements of line parameters only for the
narrow H$\beta$ and [\ion{O}{3}] lines in the whole sample (302 AGNs),  we accepted a criteria of R=log([\ion{O}{3}]/H$\beta_{NLR})=0.5$ (horizontal dashed line in Fig. \ref{f01})} as an indicator of the predominant starburst emission contribution to the narrow emission lines. We divided our sample into two subsamples:  $R<$0.5 (91 AGNs, hereafter ``SB'', starburst  dominant){ \footnote{One spectrum (SDSS J130357.42+103313.50) is excluded from the sample because  the [\ion{O}{3}] emission is comparable to  the level of  noise.}} and  $R>$0.5 (210 AGNs, hereafter ``AGN'', AGN dominant).  We should note here that such a high number of SBs is probably caused by the selection effect, since one of the criteria was that equivalent widths of typical absorption lines are small \citep[for more details see in][]{kov10}.

One could expect that where there is a significant starburst contribution, the widths of  the [\ion{O}{3}] lines would depend on the luminosity \citep{br88}. We therefore plot the log($L_{5100}$) against the log(FWHM [\ion{O}{3}]) (see Fig. \ref{f01a}) and find a good correlation between the continuum luminosity and FWHM of the [\ion{O}{3}] lines in the SB sample $R<0.5$ (r = 0.70, P = 7.3E-15), while in the AGN sample  there is no correlation. This indicates that even for the roughly estimated fluxes of the narrow H$\beta$ component there is some difference between these two sub-samples. In Fig. \ref{f4} we plot the FWHM of the broad H$\beta$ against the FWHM of the [\ion{O}{3}] for all data, where the SB sub-sample  is denoted with the full circles. As one can see, there is a good correlation between the widths for the SB sub-sample  (r = 0.74, P $<$ 1E-16).

 It is a hard task to separate AGNs, HII regions and LINERs, even if one deals with only narrow emission line objects \citep[see e.g.][]{ro97,de00}.
Therefore, the construction of BPT diagrams of broad line objects, where the permitted lines are composed of two or more components, is very difficult. \citet{na01} constructed the diagrams for a group of broad and narrow line Seyfert galaxies, and found that some Sy1 may be located in H II regions (see their Fig. 4), but there are no significant trends in the forbidden line ratios between NLSy1, Sy1 and Sy2 AGNs.  Forbidden lines can be more precisely fitted than permitted lines, but even though forbidden line ratios depend on physical properties (electron density and temperature) it is hard to separate the objects using only forbidden narrow lines.

Estimation of the narrow H$\beta$ and H$\alpha$ components (in the case of a broad line AGN) is a very complex task. There are several  problems, and one is that in principle the line profiles (especially of the  broad lines) are not pure  Gaussians, i.e., due to complex kinematics in the NLR  \citep[e.g., blue asymmetry produced by outflows, see][]{sm07} a single Gaussian very often cannot represent the shape of the narrow H$\beta$ and [\ion{O}{3}] components. Consequently the approximation of two Gaussian functions could be used, which gives some indication about velocities of the subregions. Another problem in the estimation of the narrow H$\beta$ contribution could be that the solution is non unique, i.e., a combination of different Gaussians can give different contributions of the narrow H$\beta$ component. Therefore when fitting  with multi-Gaussian components we are able  only to estimate very roughly the contribution of the H$\beta$ NLR component from the region emitting the [\ion{O}{3}] lines. We  estimated errors using a $\chi^2$ test (see Appendix I) and found that for the SB subsample, around 25\% of the AGNs (those between estimated error-bars, $R+\Delta R>$0.5) could belong to the AGN subgroup, but when we exclude these 22 objects, the correlations we obtained were not significantly changed.

We also searched for another method to separate these two groups.  We investigated the line width ratios of [\ion{O}{3}]/H$\beta_{\rm broad}$ as a function of the continuum luminosity and found that there is no correlation (see Fig. \ref{f3}), but
points with $R<0.5$ (solid circles), mostly have $R_1$=log(FWHM [\ion{O}{3}]/FWHM $H\beta_{broad})>$-0.8. Consequently, we also divided our sample into
two sub-samples with $R_1<-0.8$ (hereafter the ``AGN1'' subsample) and $R_1>-0.8$ (hereafter the ``SB1'' subsample).

\section{Results and discussion}

We calculated Spearman correlations between the spectral properties for the  total sample of 302 AGN spectra, and for the ``SB'' and ``AGN'' sub-samples with  $R<$0.5  and  $R>$0.5 respectively (which correspond closely to $R_1<$-0.8  and  $R_1>$-0.8). The results are given in Tables \ref{t01} -- \ref{t04} and Figs. \ref{f02} -- \ref{f67}.

One of the most interesting correlations is between the  FWHM (FWM10\%) of the broad H$\beta$ vs. continuum luminosity ($\lambda\mathrm{L}_{5100}$) {\bf (a similar correlation was found by \cite{st81} for the FWZI of H$\alpha$)} and the FWHM (FWM10\%) H$\beta$ vs. EW \ion{Fe}{2}.  These correlations are quite different for the SB and AGN subsamples.

The velocity-luminosity relation in AGNs has been discussed earlier. \cite{s84}  found that the FWHM of the Balmer lines increases
with luminosity for a sample of 25 AGNs. Also, \cite{wy85} reported about a correlation (r=0.5) between the H$\beta$ full width at zero intensity (FWZI) and the 4000 \AA\ continuum luminosity for a literature compilation of 94 AGNs. \cite{s99} found  a weak correlation between the width of  H$\beta$ and luminosity in the case of 126 AGNs. The same trend, i.e. a weak correlation, was found by \cite{j85}, but \cite{bg92} found an anticorrelation (r=-0.275) between the H$\beta$ FWHM and absolute V magnitude, significant at the 99\% confidence level, in a sample of 87 PG quasars. They also noted that the FWZI of H$\beta$ is quite sensitive to noise and depends strongly on the quality of the Fe II subtraction.  We therefore measured FWM10\%

In Fig. \ref{f02} we plot log($\lambda$L$_{5100}$) vs. { log(FWHM H$\beta$) and log(FWMI 10\% H$\beta$) of the broad H$\beta$ component}. {\  We found a strong correlation between the continuum luminosity and FWHM H$\beta$ for the SB and SB1  subsamples, while in the AGN and AGN1 subsamples (see Figs. \ref{f02} -- \ref{f67}), the correlation coefficients are much lower  and  not statistically significant.  The correlation  between continuum luminosity and FWMI 10\% H$\beta$ is present in the SB and SB1 subsamples, but not in the AGN and AGN1 subsamples (see Tables \ref{t01} and \ref{t02}).}

{ The relationship between the EW \ion{Fe}{2} and FWHM (FWM10\%) H$\beta$ is also different for the SBs  and AGNs  subsamples (see Fig. \ref{f03}, Table \ref{t01}). Namely, the well known anti-correlation of EW \ion{Fe}{2} and FWHM H$\beta$, which is part of the BG92 EV1 correlations, is noticeable only for the AGN and AGN1  subsamples, while the SB and SB1 subsamples show an opposite but statistically insignificant trend. Note here that \citet{g04} also found a positive correlation between the EW \ion{Fe}{2} and FWHM H$\beta$ for NLS1 objects, and a negative correlation  for broad line Sy1s.}

Some other correlations shown in Table \ref{t01} are also sample dependent: i.e., there is a significant difference in the corrlation coefficients for the AGN and AGN1 than for the SB and SB1 subsamples. As can be seen, the Baldwin effect for [\ion{O}{3}] and H$\beta$ NLR tends to be stronger in AGN and AGN1 subsamples than for the SB and SB1 subsamples.

\subsubsection{Principal component analysis}

We applied principal-component (PC) analysis for the  total sample, and subsamples.  The results of the PC analysis for the two subsamples and for the total sample are shown in Tables \ref{t03} and \ref{t04}. As can be seen, { the AGN dominant} subsamples (AGN and AGN1) may be described with the first three eigenvectors which represent $\sim$ 82\% of variance (see Table \ref{t03} and \ref{t04}). The first eigenvector, which describes $\sim 39$\% of variance, represents dependence of  line EWs vs. continuum luminosity (i.e., the Baldwin effect).  The EW \ion{Fe}{2} vs. EW [\ion{O}{3}] anticorrelation is clearly present, implying that it is related to the Baldwin effect. Note, that in this eigenvector there is no influence from the FWHM H$\beta$. The second eigenvector  represents a relationship between line EWs and FWHMH$\beta$, which should be similar to the BG92 EV1, but in our case, the EW \ion{Fe}{2} vs. EW [\ion{O}{3}] anticorrelation is not detected. In our second eigenvector we found the anticorrelation of FWHM H$\beta$ vs. EW \ion{Fe}{2} and the expected correlation between FWHM H$\beta$ and EW H$\beta$ {\  broad}. Our third eigenvector  is dominated by  the  broad H$\beta$ component and EW \ion{Fe}{2} indicating a connection between the BLR and \ion{Fe}{2} emitting region \citep[see][]{kov10}.

The PC analysis results for the SB and SB1 subsamples are significantly different(Table \ref{t03} and \ref{t04}).  THe variance may also be described with the first three eigenvectors representing 84\% (SB) and 76.5\% (SB1) of variance. The first eigenvector accounts for 48.5\% and 41.4\% respectively of the variance of the subsamples, which implies that it dominates in both SB subsamples.
{ It represents a relationship between the line EWs and continuum luminosity (the Baldwin effect), but also correlations/anticorrelation between the EWs  with the broad H$\beta$ FWHM, since a strong correlation between FWHM H$\beta$ and $\lambda\mathrm{L}_{5100}$ is present. The EW [\ion{O}{3}] vs. EW \ion{Fe}{2} anticorrelation is present also as the Baldwin effect of the lines. This eigenvector does not represent the BG92 EV1 since the FWHM H$\beta$ and EW \ion{Fe}{2} have a positive correlation.

\subsection{Possible physical interpretation of results}

The most interesting result is that SBs subgroups show correlations between the narrow and broad line widths and luminosities while the AGN subgroups do not.
It has previously been noticed that some correlations between spectral properties depend on the FWHM H$\beta$  range of the observed sample  \citep[see][etc.]{st81,g04,su09,za10,kov10}. For example, \citet{g04} found that the FWHM H$\beta$ and EW \ion{Fe}{2} anti-correlate for AGNs with broad lines (BLSy1s), but anti-correlate for NLSy1s. Here we found two subsamples, one where there is a significant correlation between the FWHM H$\beta$ broad component and continuum luminosity, and another where this correlation is not present.
It is interesting that the correlation between continuum luminosity and FWM10\% H$\beta$  is  the same as  for the FWHM H$\beta$ (see the SB and SB1 subsamples in Tables \ref{t01} and \ref{t02}) .

Among other things, the different behavior of the two groups may be caused by: a) different contributions of starburst regions, i.e., a different nature of the ionization source, or b) by different Eddington ratios and/or  geometrical
structures of the BLR.

\subsubsection{A starburst contribution -- a difference in the source of ionization?}

One can see from  Tables \ref{t01} and \ref{t02} that { there are significant} differences in spectral properties between the two subsamples.   A possible explanation of these differences may be that there is a difference in the source of ionization.

As we mentioned above, there  is a possibility that  an AGN spectrum in the earlier activity phase  is composed { of a  starburst  and a central engine (pure AGN) spectrum }\citep[see][]{cr02,WW06,WW08,m09}. Moreover, recently \citet{pop09} found that in the case of NLSy1 galaxy Mrk 493, the narrow-line ratios correspond to starbursts rather than to an AGN origin.

In the case of a dominant contribution from a starburst region to emission in the narrow lines,  one can expect that, besides the AGN emission, additional  significant  starburst  emission may be present in the composite continuum. Then, there is probably an extensive source of continuum emission \citep[as in the case of Mrk 493, see][]{pop09} that may be produced by a number of exploding stars.  Consequently, the luminosity of such objects will depend on the star  forming rate. In { the starburst-dominant subsamples there is a positive, but statistically insignificant trend between the FWHM H$\beta$ and} EW \ion{Fe}{2}, and it seems that in general the \ion{Fe}{2} emission is stronger in this subsample (log(EW FeII)$\sim$1.7-2.3). The strong \ion{Fe}{2}
emission and weak [\ion{O}{3}] emission can be explained with a model that
contains a massive starburst (SB) plus an AGN \citep[see e.g.][]{lt06}.
The SB+AGN can lead to large scale expanding
supergiant shells. A good correlation between the FWHM and luminosity in one group of AGNs may indicate that  the broad component is not primarily a function of geometry (e.g., by rotation due to the gravitational force), but rather by random motion of the gas caused by different effects (such as the gas being randomly accelerated in several bursts). { In such a case part of the flux of the broad Balmer emission lines may arise in the stellar envelopes of Wolf-Rayet and OB stars  associated with multiple SN events \citep[see][and references therein]{Iz07}. Note that EW H$\beta$ broad only increases with $\lambda\mathrm{L}_{5100}$ in the SB and SB1 subsamples, and this increase is not found in the AGN subsamples.}

One can speculate that the FWHM of the broad component might not be formed as in the classical broad line region (where rotation velocity should be present), but as a sum of random motions of emitters \citep[as e.g. in extended envelopes, see][]{n06}. If this is the case,   it
might be that \ion{Fe}{2} lines are also formed in the gas located in (or
around) the starburst regions, i.e., the intensity of Fe II lines does not depend on the geometry of the BLR. The star forming regions can affect observed spectra of AGNs. As an example,
recall here the results obtained by \citet{cr02}, who found a strong anti-correlation with luminosity for the equivalent widths of [\ion{O}{2}]$\lambda$3727 and who suggested that the [\ion{O}{2}] line observed in the high-luminosity AGNs may be emitted, to a large part, by intense star-forming regions; i.e., the AGN contribution to this line could be weaker than previously assumed.

On the other hand, {analysis of} the AGN and AGN1 samples shows that there is no correlation between the widths (kinematics) of broad H$\beta$ and luminosity. This would be expected if the widths were caused by rotational component of motion (by geometry), as is assumed in the classical BLR. Additionally, the EW \ion{Fe}{2} shows a trend of being weaker as widths increase. This also can be caused by geometry, or it could be that as we  look deeper into the BLR, the \ion{Fe}{2} emission becomes weaker.

In the case of a pure (or dominant) AGN emission  there is a point-like photoionization source that  influences  the NLR emission: as the central source gets stronger, the continuum { also becomes stronger, and it } affects the processes in the NLR, as well as the size of the region. The \ion{Fe}{2} emission becomes stronger.  This may be caused by additional atomic processes in some parts of the BLR. In this case, the BLR is formed around the central black hole, { and its geometry and gravitational motion influences line profiles (widths)}. Since dimensions will be affected by the central source luminosity there will be correlations between the broad line EWs and luminosity. However, the BLR geometry can be quite different and one cannot expect correlations between the broad line FWHM (FWM10\%) and luminosity. It is clear that in this case, the Baldwin effect and BG92 EV 1 are caused by the luminosity of  central source and line-forming processes in the NLR and part of the BLR \citep[probably in the ILR, see][]{kov10}.

\subsubsection{Accretion  rate}

The ratio between the bolometric and Eddington luminosities can be taken as equal to the dimensionless accretion rate $\dot{m}\sim L/M$ and
it is well known that the Boroson and Green EV1 is closely related to the Eddington ratio, L/M  \citep[see e.g.][]{ma01,b02}.
The mass of a black hole ($M$) can be estimated using  \citep{wy85,pr88,vp06,mc08}

$$M \sim L^{0.5}_{5100}(FWHM)^2,$$
where FWHM is the full width at half maximum of the broad component. From this, one can extract an  expected relationship between the luminosity, accretion rate and FWHM as

$${\rm log}(L)={\rm const}+ 2\cdot {\rm log}(L/M) + 4\cdot {\rm log}(FWHM).$$
Consequently, if two groups of AGNs (in our case SB and SB1) have similar Eddington ratios, there will be  a high correlation between the luminosity and FWHM. It may be an explanation for a good correlation between luminosity and FWHM seen in the SB and SB1, but the question remains, why in the AGN and AGN1 subsamples is there no correlation? Also, why is this correlation connected with the ratio of the narrow lines?  One solution might be that SB and SB1 are young objects where starburst regions also contribute to the line emission (especially to the narrow lines), and that they have a similar accretion rate.  In that case we can expect a good correlation between the L and FWHM, and dominant starburst emission in narrow lines.

\section{Conclusions}

In this paper we have investigated connections between the Baldwin effect and the BG92 EV1 and different correlations between spectral characteristics of a sample of 302 AGNs; dividing the sample with respect to the narrow [\ion{O}{3}] and H$\beta$ ratio and also by using the ratio of widths of [\ion{O}{3}] and H$\beta$.

We have investigated different correlations within the  subsamples.  { We} found that:

{
i) It seems that the line width of broad line Sy1s introduced by some authors  \citep[FWHM=3000 km/s or 4000 km/s, see][etc.]{g04,su09,za10,kov10} is not relevant for separation of AGNs with broad lines into two groups (BLS1 and NLS1, see Fig. \ref{f3}), since some AGNs with FWHM$<$ 3000 km/s or $<$ 4000 km/s have the same characteristics as AGNs with broader lines. We propose to divide the subsamples of AGNs using as a criterion the flux ratio of the narrow [\ion{O}{3}] and narrow H$\beta$ lines, or by using the ratio of the FWHMs of [\ion{O}{3}] and broad H$\beta$, or H$\beta$ narrow and H$\beta$ broad components.\footnote{Since the width of the H$\beta$ narrow and the corresponding  [\ion{O}{3}] line component is essentially the same}}

ii) There is a significant difference in the correlations between line properties for the two subsamples.  This indicates that there is a significant difference in the physics of emitting gas and the origin of the broad line components. We found that BG92 EV1 and the Baldwin effect have the same physical background  in both subsamples. In  the SB and SB1 subsamples, eigenvector 1  shows strong acorrelation between luminosity, the { FWHM of broad H$\beta$}, and the EWs of lines, while in the other subsamples there are high correlations only between luminosity and EWs of lines.

iii) The narrow and broad line widths  of the SB and SB1 subsamples are luminosity dependent and they seem not to be connected with only predominant rotational motion, but rather by randomly distributed high velocity gas. Conversely, in the case of the AGN and AGN1 subsamples, there is no  FWHM$_{H\beta}$ broad line dependence on luminosity and the widths seem to be influenced by the BLR geometry. Such a high correlation between the luminosity and FWHM in the SB and SB1 subsamples may be explained as: a) a connection between starburst regions and  the region where the broad H$\beta$ line is formed, and b) the SB and SB1 subgroups having very similar accretion rates, in which case one can expect a high correlation between the luminosity and FWHM.

\acknowledgments

This work is a part of the project  ``Astrophysical
Spectroscopy of Extragalactic Objects'' supported by the Ministry of Education and
Science of Serbia. The authors would like to thank the Alexander von Humboldt (AvH) foundation for their support of this work through the grant "Probing the Structure and Physics of the BLR using AGN Variability" within the AvH program for funding research group linkage. { We would like to
thank to Todd Boroson and Martin Gaskell for very useful comments.}

\clearpage

\appendix

\section{Estimation of the narrow H$\beta$ component}

As we mentioned above, estimating the narrow H$\beta$  component is a problem, especially in the case of broad-line AGNs. In principle, due to complex kinematics in the NLR the narrow lines usually cannot be represented by a pure  Gaussian profile. But, if one would like to estimate the contribution of the NLR to the whole line, a Gaussian profile approximation may be used.  This also gives some indication about velocities in the subregions as well as the contribution of the NLR to the total H$\beta$ flux. Furthermore there is the problem of non-uniqueness in the solutions, i.e., that combinations of different Gaussians can give different contributions to the total  H$\beta$ line.

In order to estimate the error-bars of the fits of the H$\beta$ NLR intensity we perform the following procedure: we started from the best fit obtained, and we considered the different deviation of the H$\beta$ NLR intensity  as $f\times {I_{H\beta NLR}}$, where ${I_{H\beta NLR}}$ is the H$\beta$ NLR intensity obtained from the best fit. The $f$ values are varied from 0.9 to 0.1 and from  1.1 to 1.9 with a step of 0.1. For each value of  $f\times {I_{H\beta NLR}}$ we again looked for the best fit and found values of $\chi^2$. We then constructed $\chi^2/\chi^2_0$ vs. $R$ diagrams\footnote{$\chi_0$ is the value obtained in the primary fit} (see Figs. \ref{fnew} and \ref{fnew1}). We  found three characteristic $\chi^2/\chi^2_0$ vs. $R$
shapes: (i) with a deep minimum, (ii) with a broad minimum, and also (iii) with an almost flat minimum in a wide interval of the narrow H$\beta$ intensity (see Figs. \ref{fnew} and \ref{fnew1}). Next we assumed that a change of 5\% of the $\chi^2$ may be taken as an indicator of the error-bar, i.e., $\chi^2/\chi_0^2=1.05$ (horizontal lines in Figs.  \ref{fnew} and \ref{fnew1}).

In additional, we inspected spectra in the H$\beta$+[OIII] spectral region, and
found that there are three characteristic spectra (as shown in Fig. \ref{fnew}): (i) where the narrow H$\beta$ is prominent and easy to
fit, (ii) where the narrow H$\beta$ is not so prominent, but can still be clearly seen and (iii) where the narrow H$\beta$ cannot be clearly seen, but the best fit indicate existence of the narrow component. As can be seen in Fig. \ref{fnew} the [O III] lines usually can be fit well with two Gaussians for each [O III] line, but the fitting procedure of the H$\beta$ NLR line could sometimes be very difficult because the narrow component of H$\beta$ cannot be distinguished well especially in the third case. For that reason, the error in the determination of the H$\beta$ NLR intensity  dominates the error of the [O III]/H$\beta$ NLR ratio. In Fig. \ref{fnew1} we present three cases of the spectra where the narrow H$\beta$ component is too weak, and, as can be seen, there are also cases where the $\chi^2$ test shows a very important contribution of the narrow H$\beta$ component.

Using this test we found 84 cases from the AGN and one case from the SB subsample, where the H$\beta$ NLR intensities changed in a broad range, but $\chi^2$ shows only very small changes, which cannot reach 5\% of $\chi^2_0$.  This means that the contribution of the narrow H$\beta$ in these objects is negligible.  We also found 26 objects in the SB subsample for which the $R=$[O III]/H$\beta$ ratio has an error-bar which indicates that it could be higher than 0.5. Finally, in the SB subsample we found 69 objects with $R+\Delta R$ clearly $<0.5$.
It is interesting that the correlations obtained are not changed when we exclude the spectra where the estimates of the narrow H$\beta$ contribution have large error-bars.

\section{Error-bars of the FWHM and FWHM10\% estimates}

As also mentioned above, the  H$\beta$  broad component was taken to be the sum of the H$\beta$ ILR and H$\beta$ VBLR components (see Fig. \ref{f09}). Since there may be a problem in subtraction of the narrow component, we did a test by  considering first the objects with flat minimum (i.e., where the narrow H$\beta$ component cannot be properly estimated). We then fixed the narrow component taking  $f=0.1$ and $f=1.9$ and found the best fit of the broad component. We measured FWHM and FWMI10\%  from these two broad components (using the ILR and VBLR) and found that the error-bars were in the range of 5\%-7\% (in all cases less than 10\%) in the case where the FWHM has a value between 1000 and 2000 km s$^{-1}$.

\clearpage

\begin{figure}
\includegraphics[width=0.5\textwidth,angle=-90]{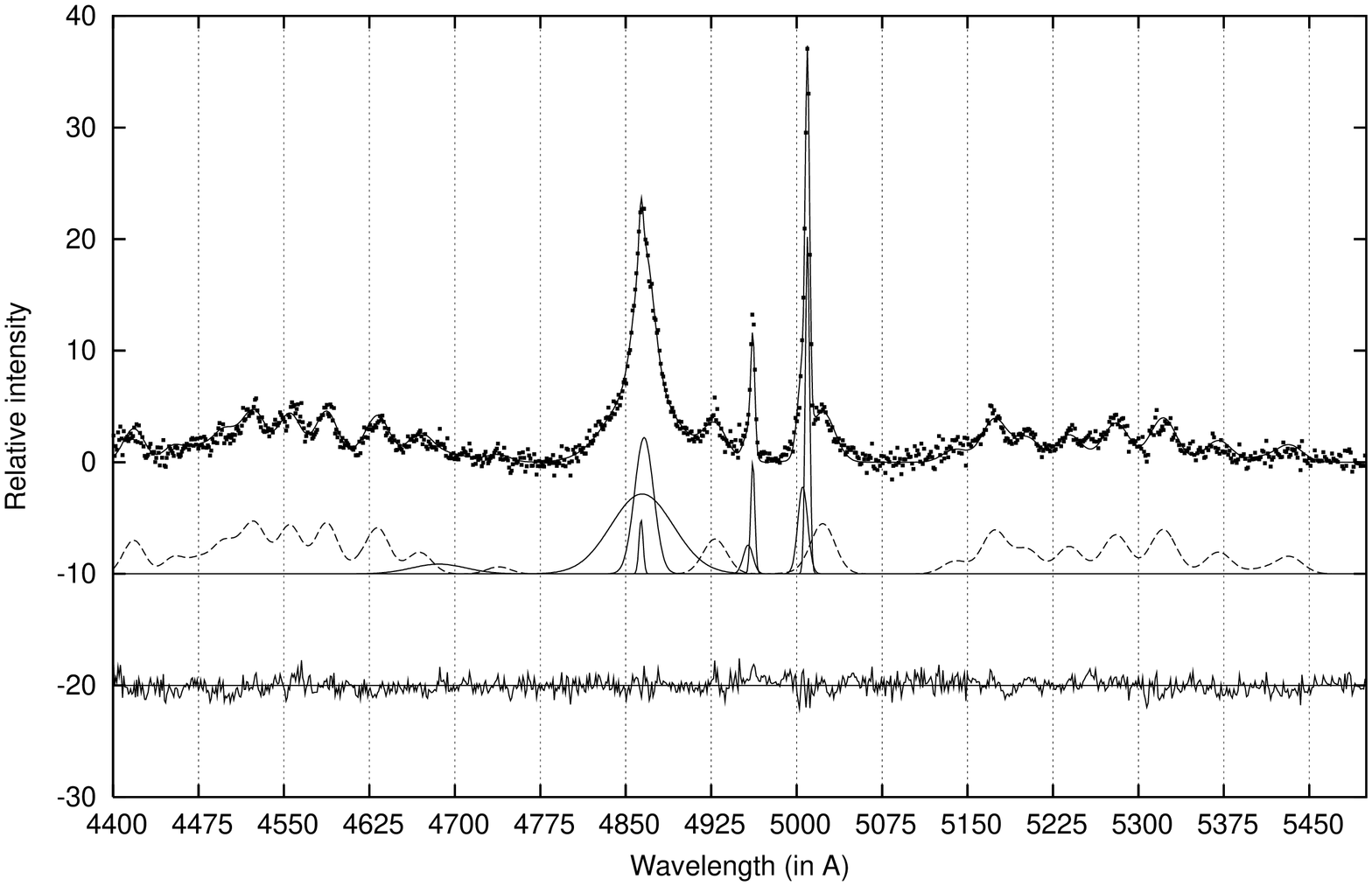}
\caption{The best fit of the H$\beta$ region (see text).}
\label{f1}
\end{figure}

\begin{figure}
\includegraphics[width=0.5\textwidth]{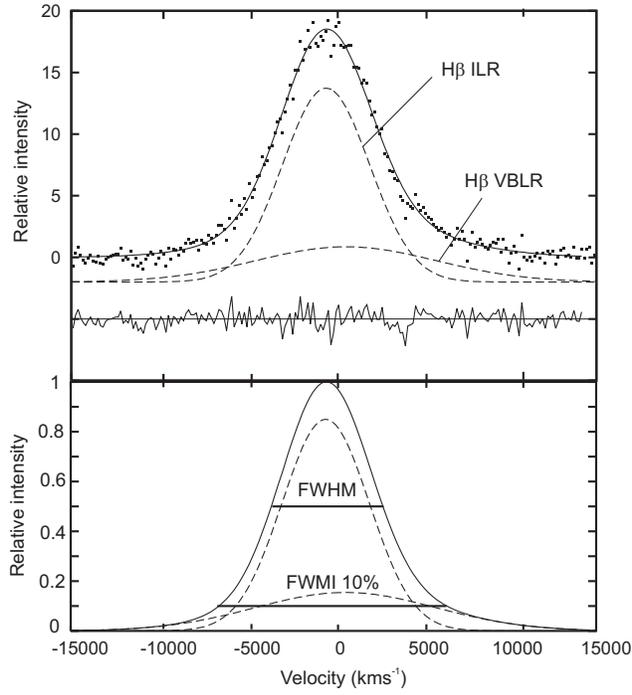}
\caption{Measurement of the full width at half maximum and at 10\% of the maximum (see text).}
\label{f09}
\end{figure}

\begin{figure}
\includegraphics[width=0.6\textwidth]{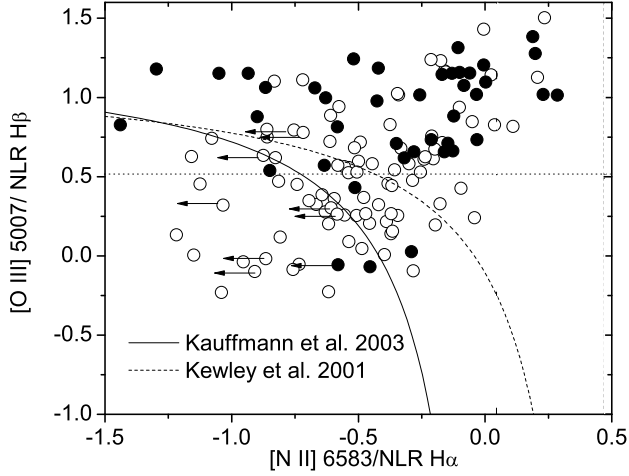}
\caption{The BPT diagram for the  subsample of 137 AGNs containing the narrow  H$\beta$ and  H$\alpha$ lines. Open circles denote AGNs with FWHM of the broad H$\beta$ smaller than 3000 km s$^{-1}$, while full circles represent AGNs with FWHM $>$ 3000 km s$^{-1}$.  Arrows indicate upper limits for the AGN where maximal [N II] emission has been estimated (since the line was too weak). The horizontal dashed line represents the assumed border between 'AGN' and 'starburst', i.e., $R=$log[\ion{O}{3}]/H$\beta_{NLR}=$0.5. The two curved lines represent the borders given in \citet{k03} and \citet{k01}.}
\label{f01}
\end{figure}

\begin{figure}
\includegraphics[width=0.35\textwidth,angle=-90]{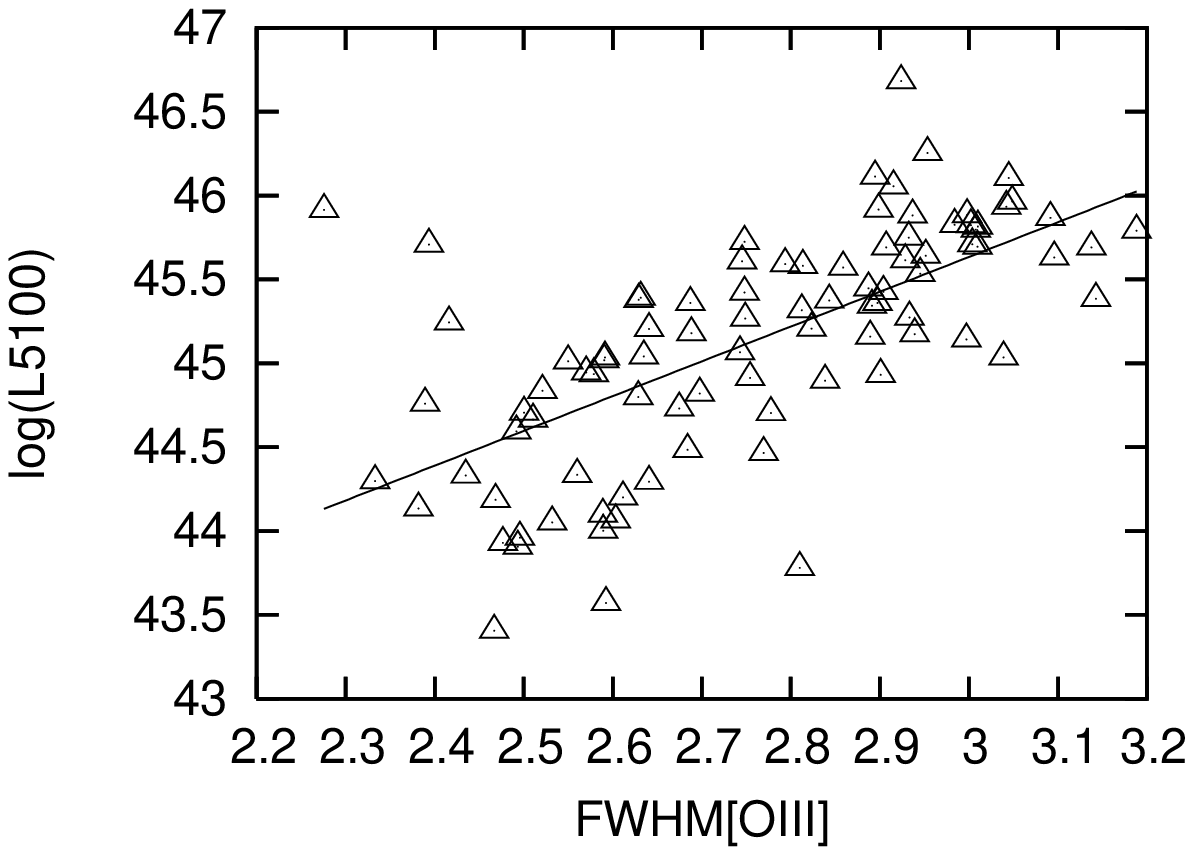}
\includegraphics[width=0.35\textwidth,angle=-90]{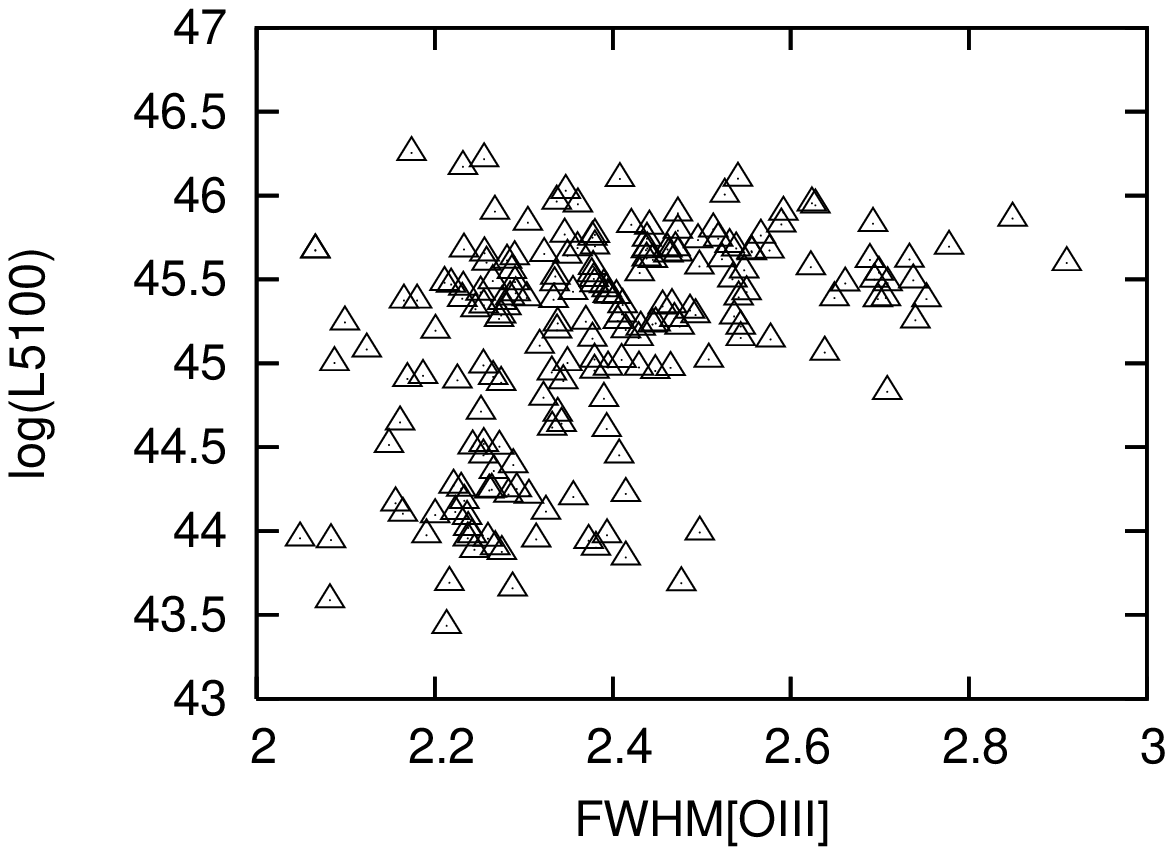}
\caption{The continuum luminosity vs. FWHM of the narrow [\ion{O}{3}] lines. The left panel is for $R<0.5$, and the right panel for $R>0.5$}
\label{f01a}
\end{figure}

\begin{figure}
\includegraphics[width=0.4\textwidth,angle=-90]{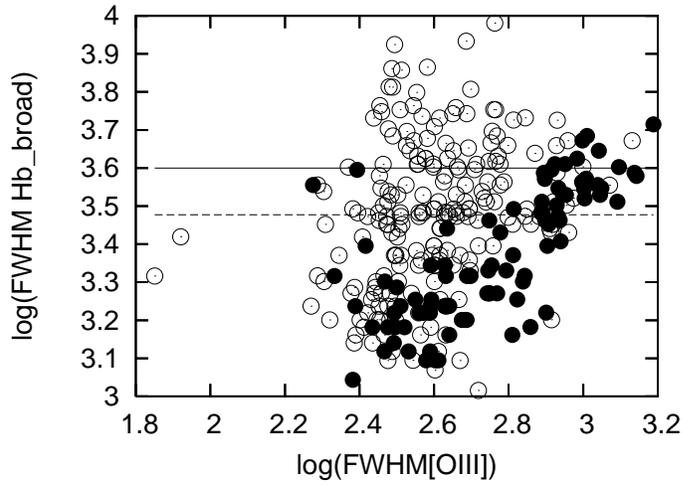}
\caption{FWHMs of the broad H$\beta$ vs. the narrow [\ion{O}{3}] lines (which corresponds to the narrow H$\beta$). Filled circles denote the sample with $R<0.5$. The solid line denotes a border of FWHM H$\beta =$ 4000 km/s \citep[see][]{ma10}, while the dashed line denotes FWHM H$_\beta =$ 3000 km/s \citep[see][]{kov10}.}
\label{f4}
\end{figure}

\begin{figure}
\includegraphics[width=0.35\textwidth,angle=-90]{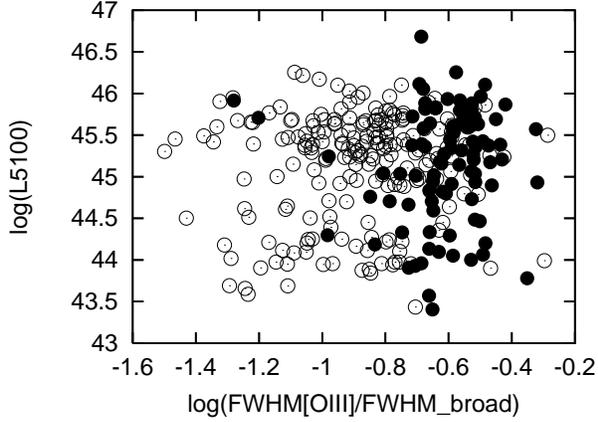}
\caption{The continuum luminosity vs. the ratio of FWHMI [\ion{O}{3}] and broad H$\beta$.}
\label{f3}
\end{figure}

\begin{figure}
\includegraphics[width=0.3\textwidth,angle=-90]{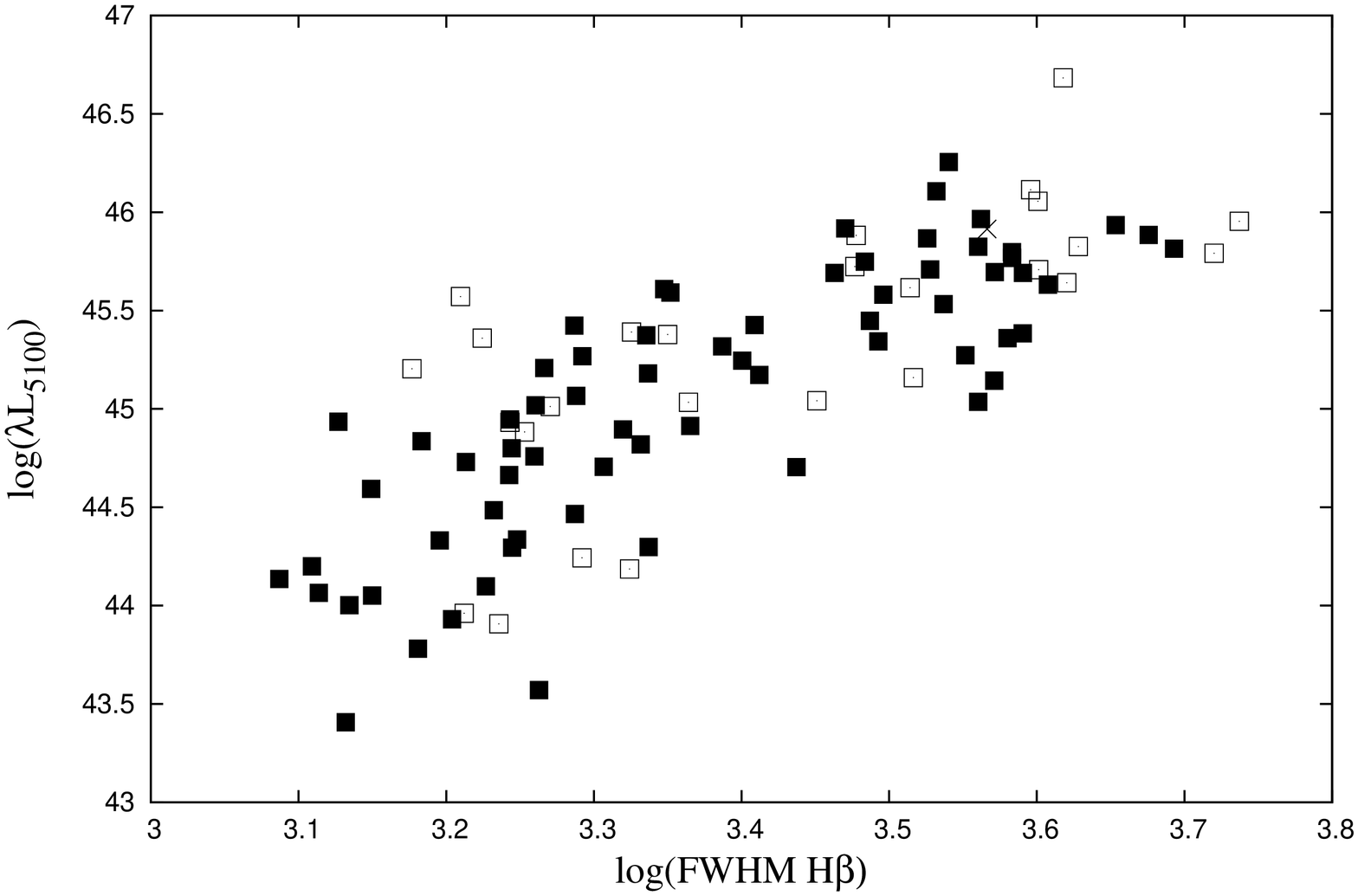}
\includegraphics[width=0.3\textwidth,angle=-90]{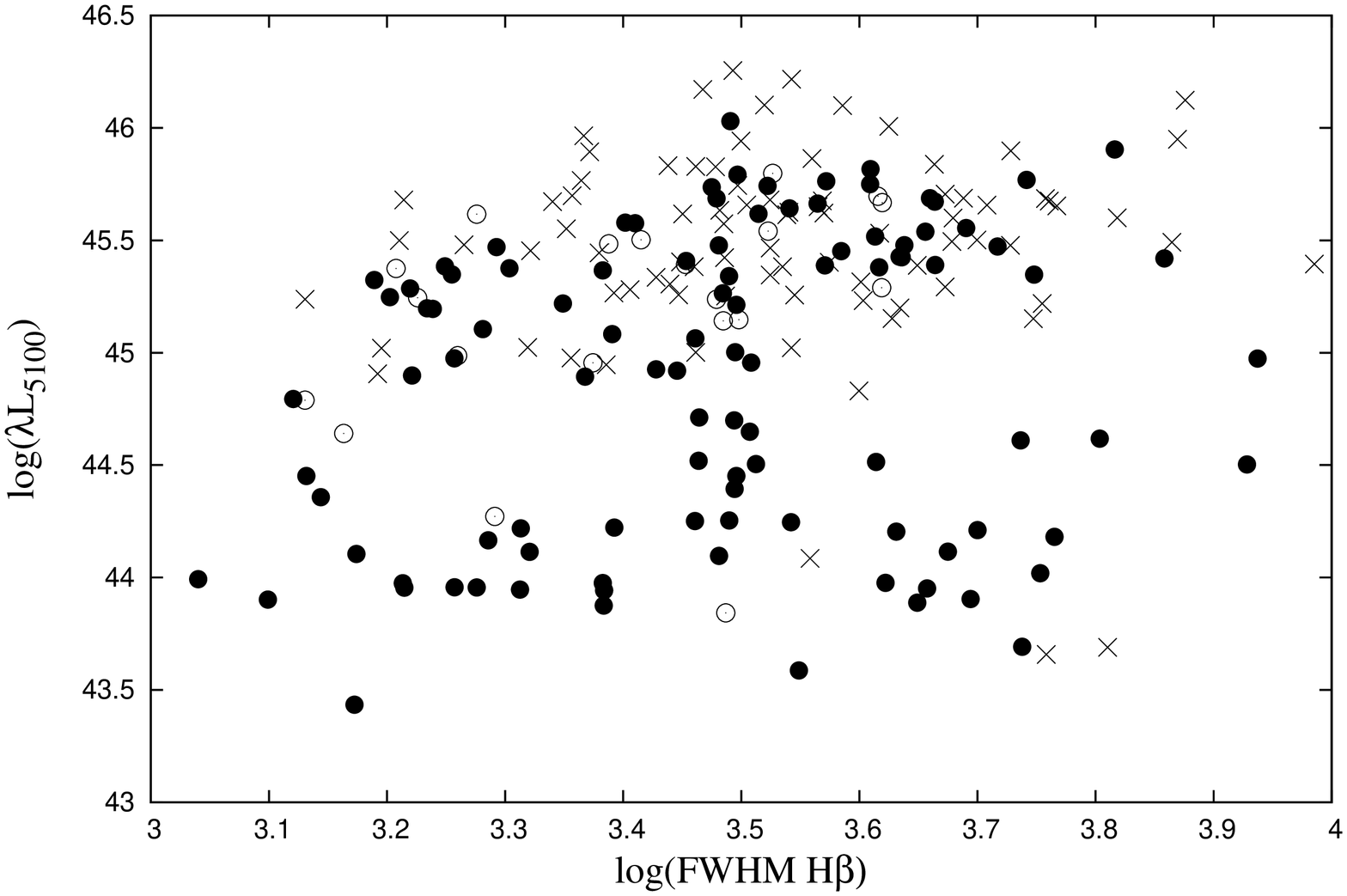}
\includegraphics[width=0.3\textwidth,angle=-90]{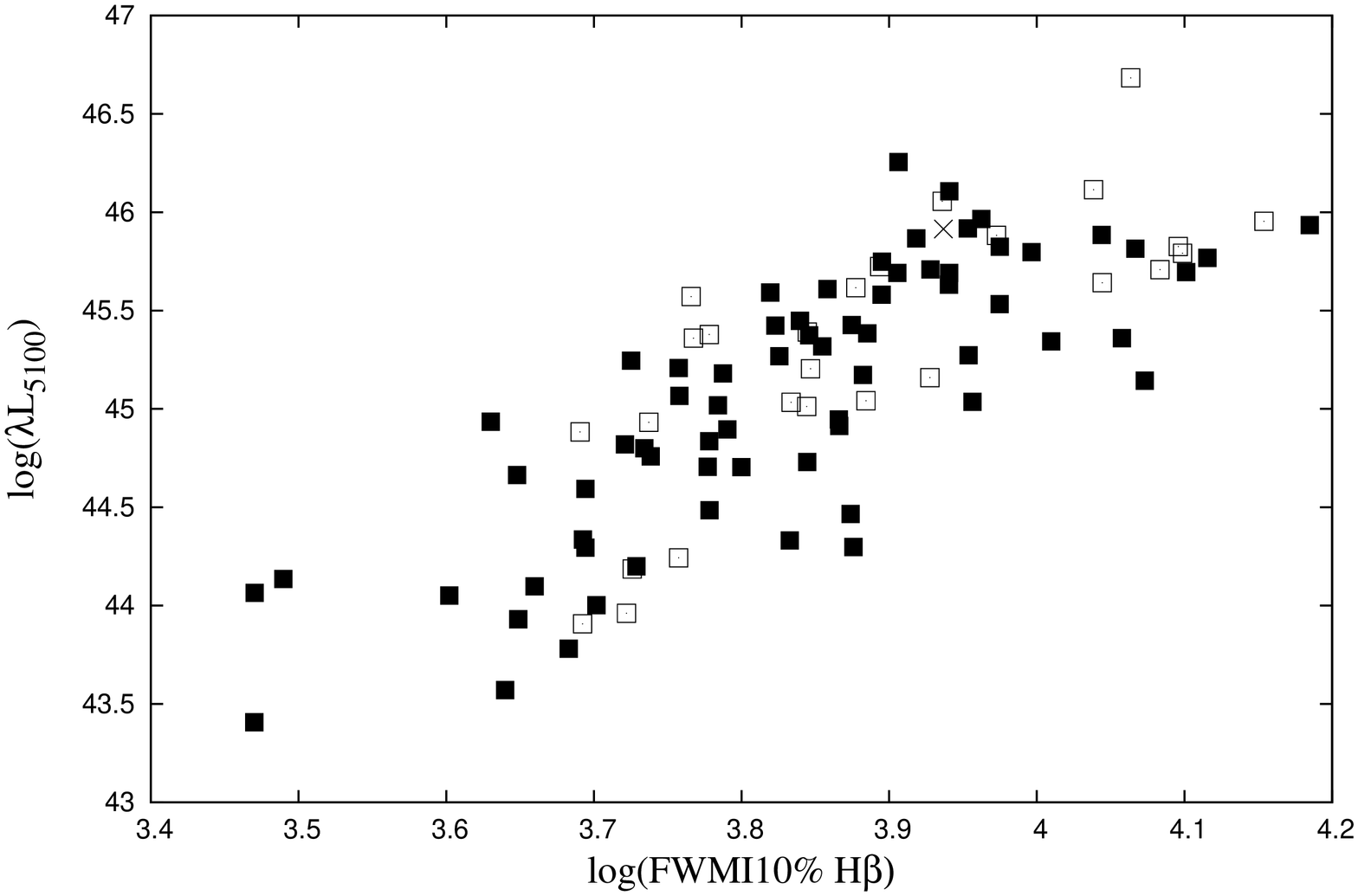}
\includegraphics[width=0.3\textwidth,angle=-90]{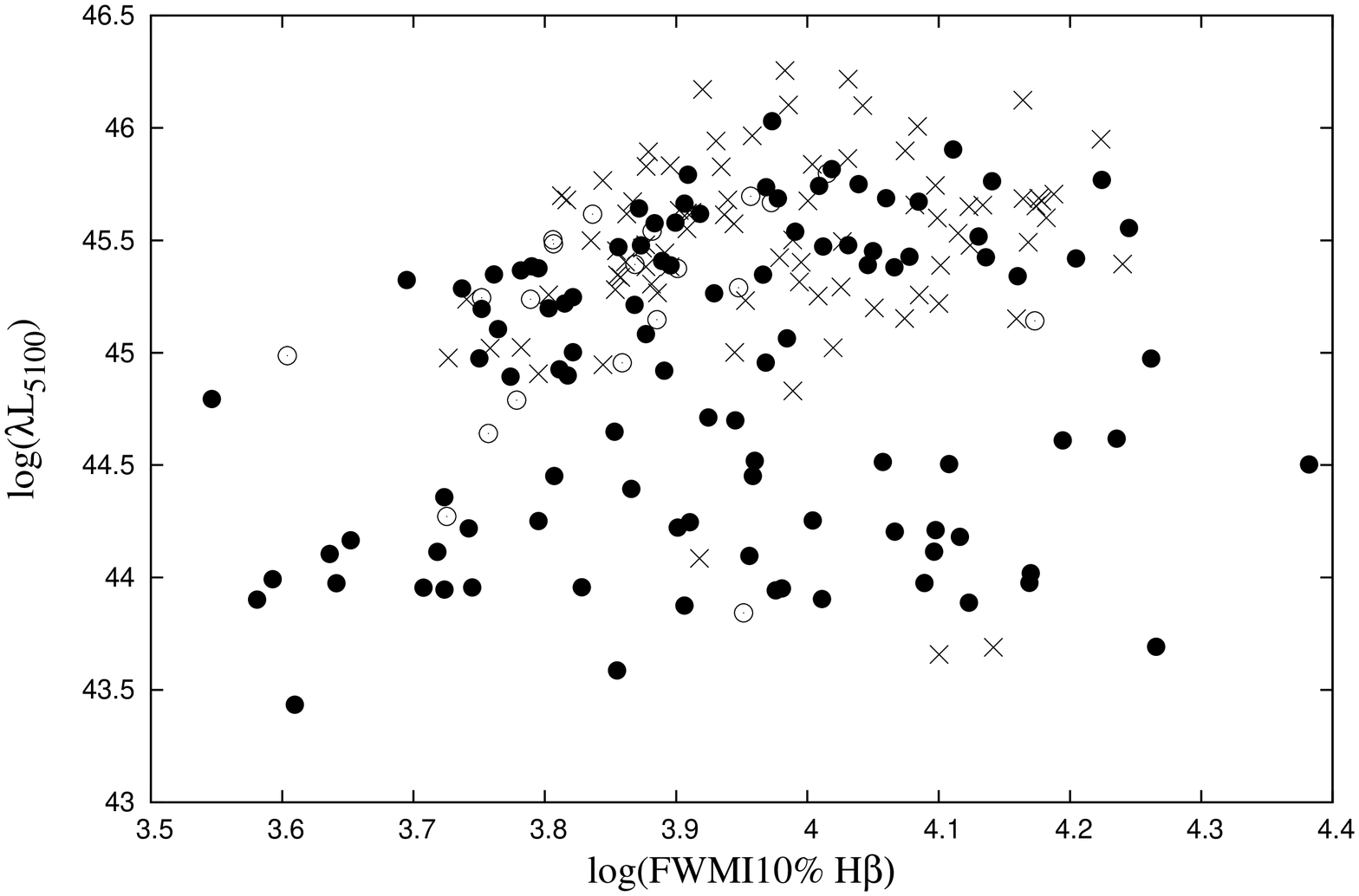}
\caption{Correlations between continuum luminosity and FWHM H$\beta$ (top panels) or FWM10\% H$\beta$ (lower panels) for AGNs with $R<$0.5 (left panels) and $R>$0.5 (right panels). Notation (see Appendix):  full squares -- $R+\Delta R<$0.5, open squares -- $R+\Delta R>$0.5,  crosses -- flat minimum, full
circles -- $R-\Delta R>$0.5, open circles --  $R-\Delta R<$0.5}
\label{f02}
\end{figure}

\begin{figure}
\includegraphics[width=0.3\textwidth,angle=-90]{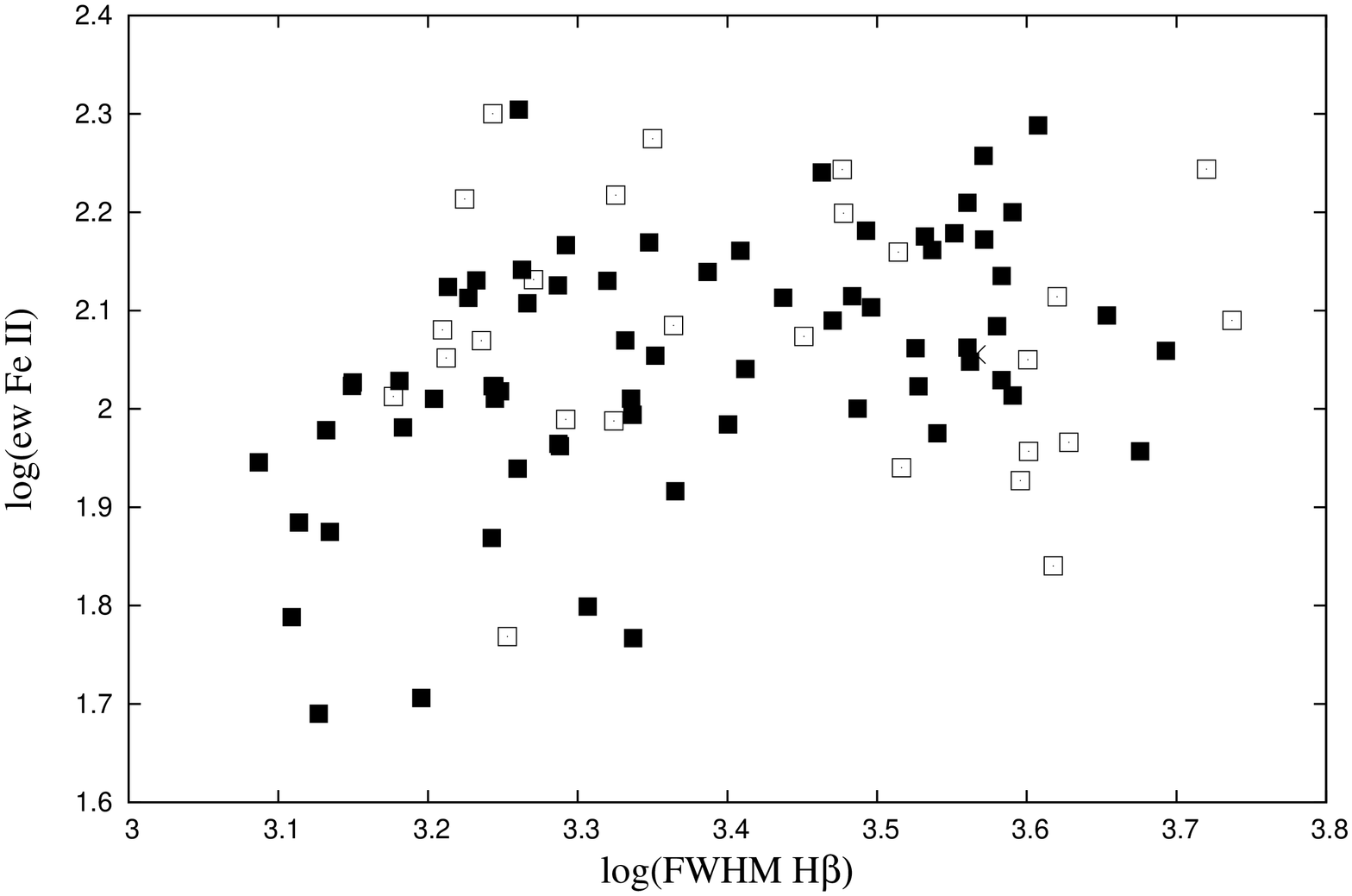}
\includegraphics[width=0.3\textwidth,angle=-90]{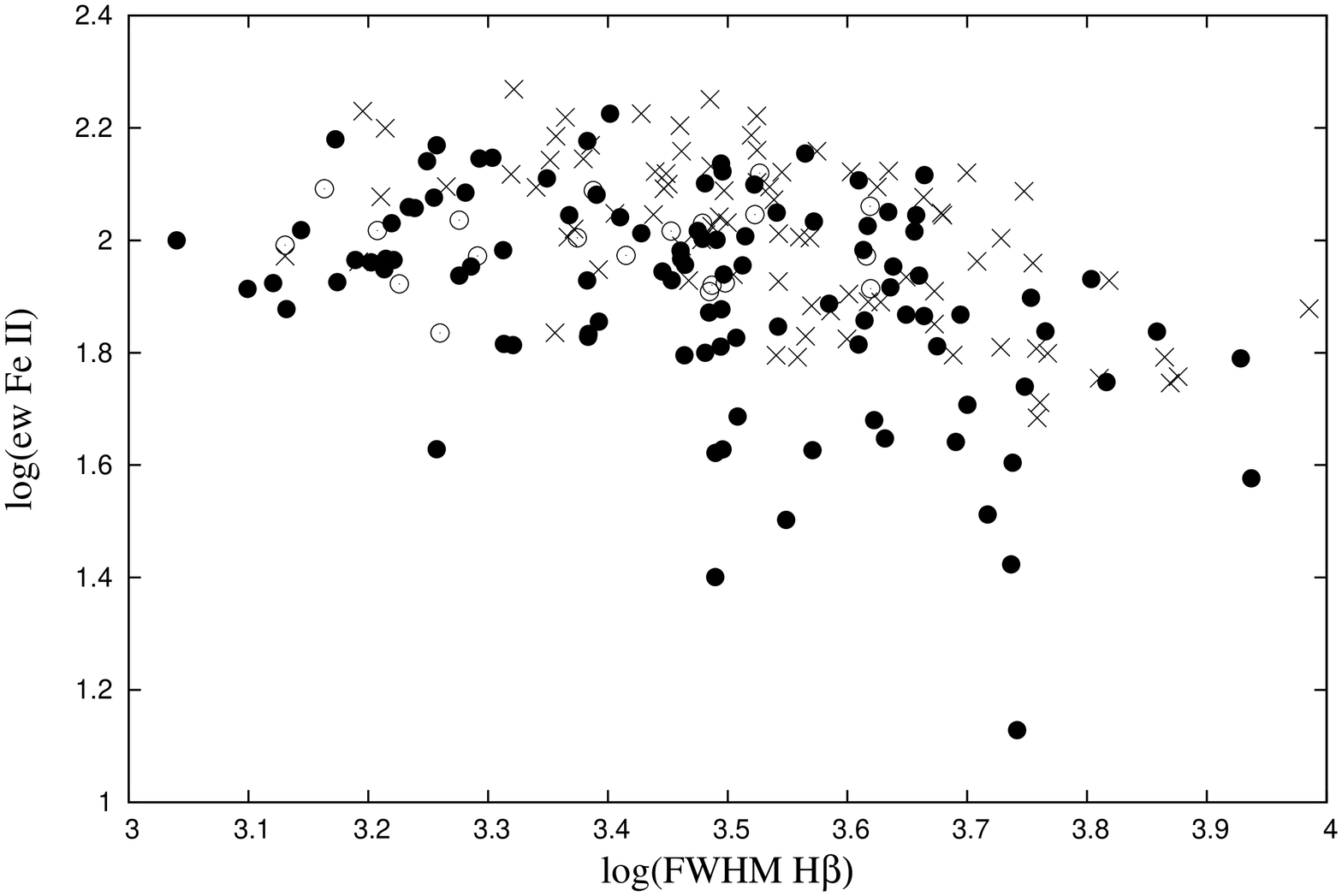}
\includegraphics[width=0.3\textwidth,angle=-90]{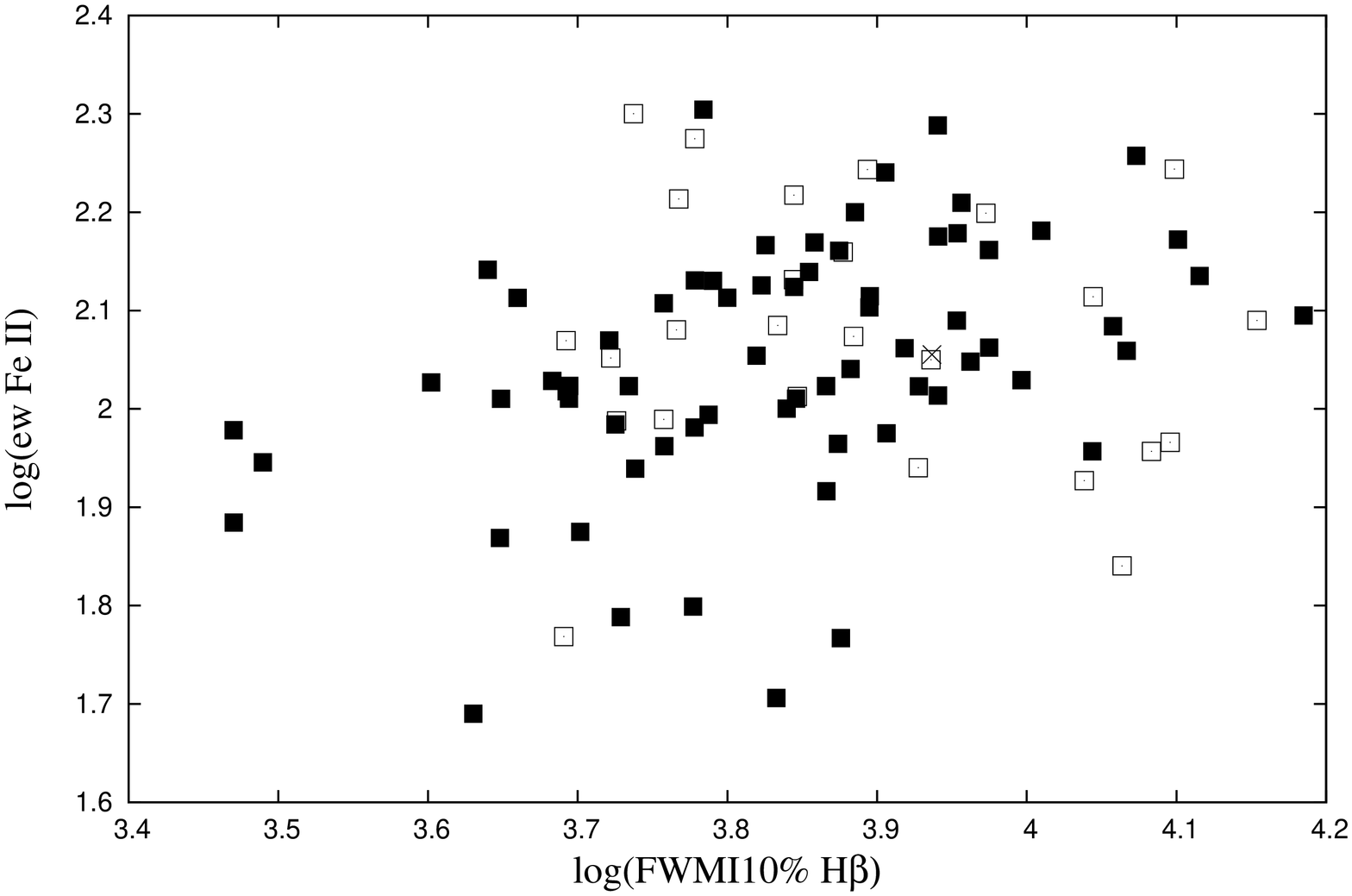}
\includegraphics[width=0.3\textwidth,angle=-90]{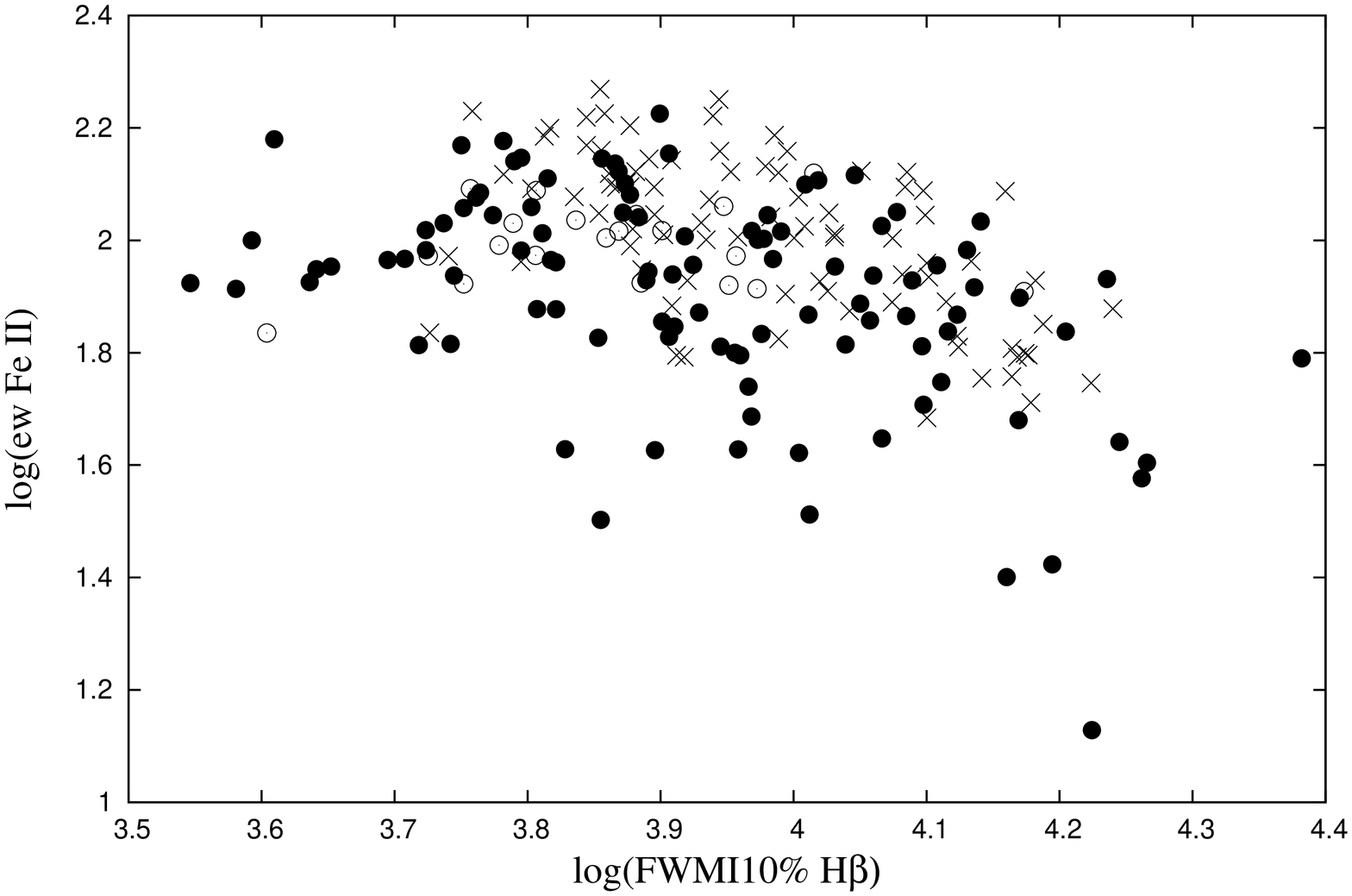}
\caption{The same as in Fig. \ref{f02} but for EW \ion{Fe}{2}.}
\label{f03}
\end{figure}

\begin{figure}
\includegraphics[width=0.37\textwidth,angle=-90]{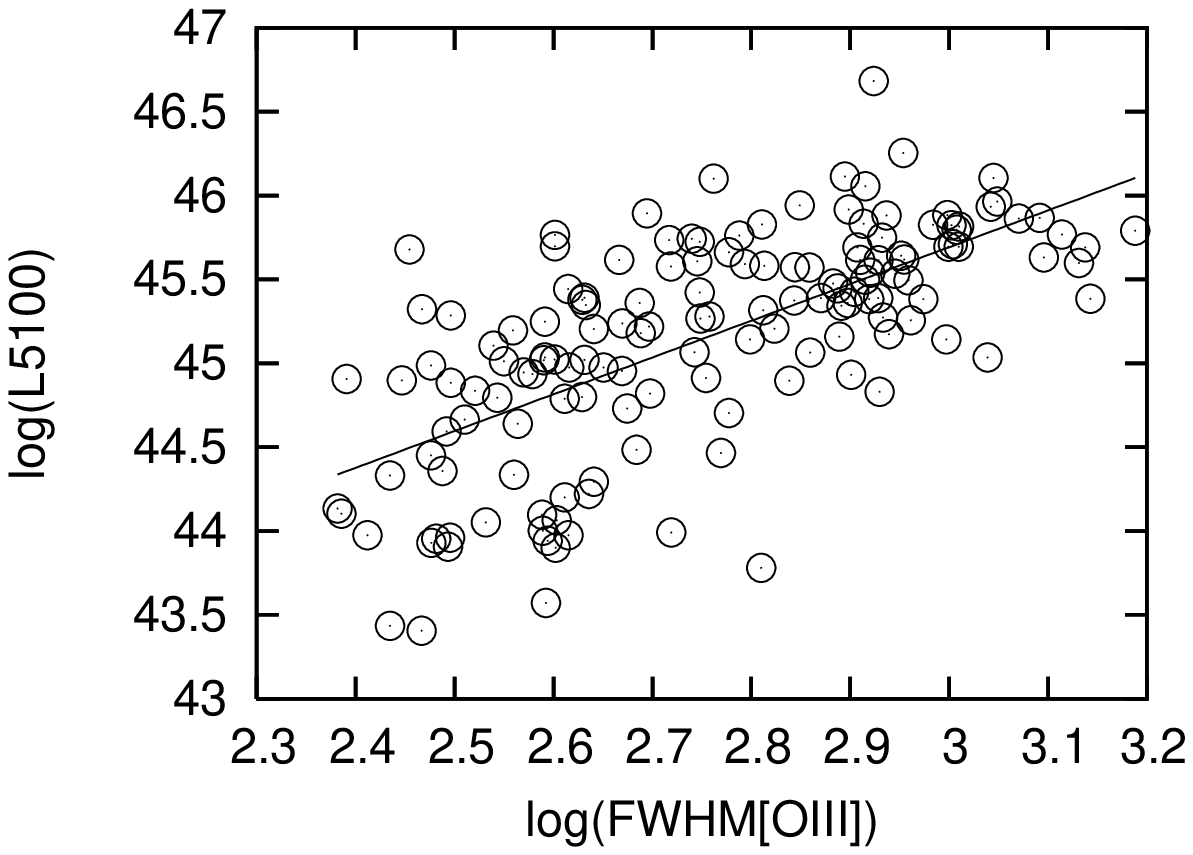}
\includegraphics[width=0.37\textwidth,angle=-90]{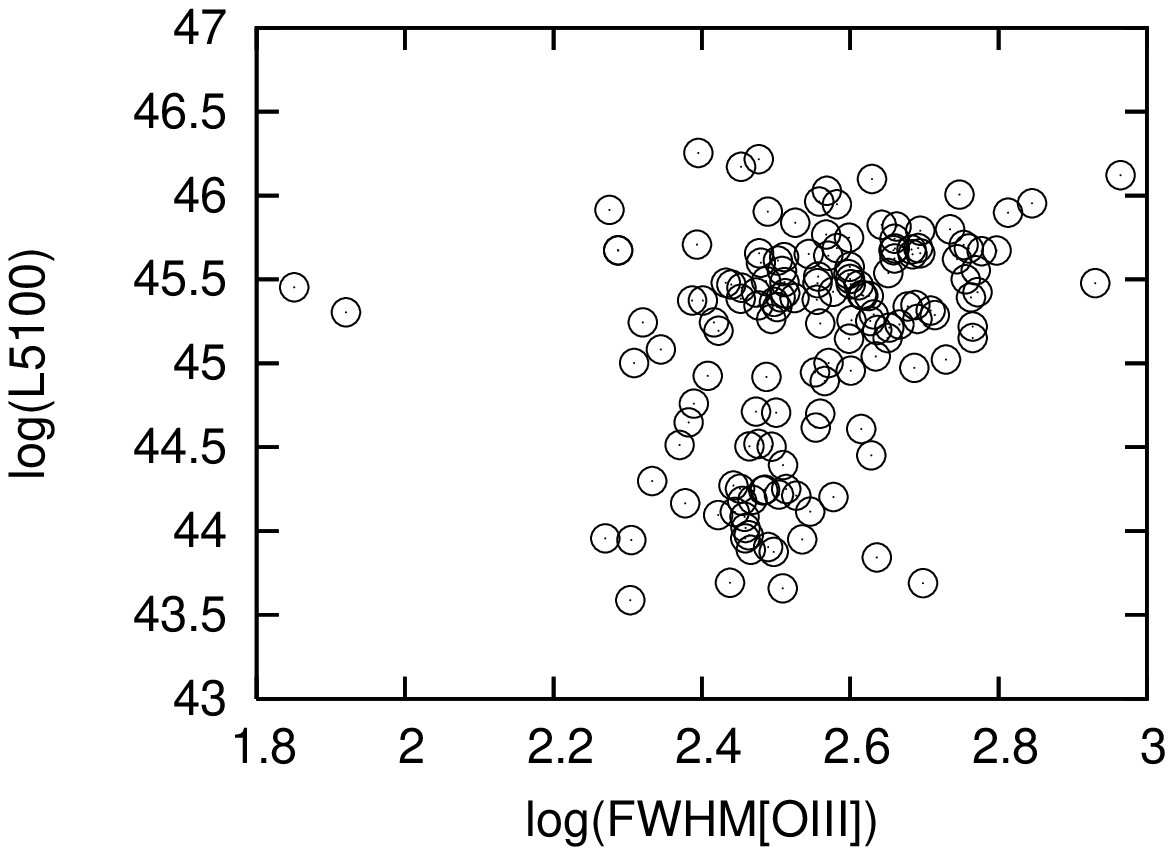}
\includegraphics[width=0.37\textwidth,angle=-90]{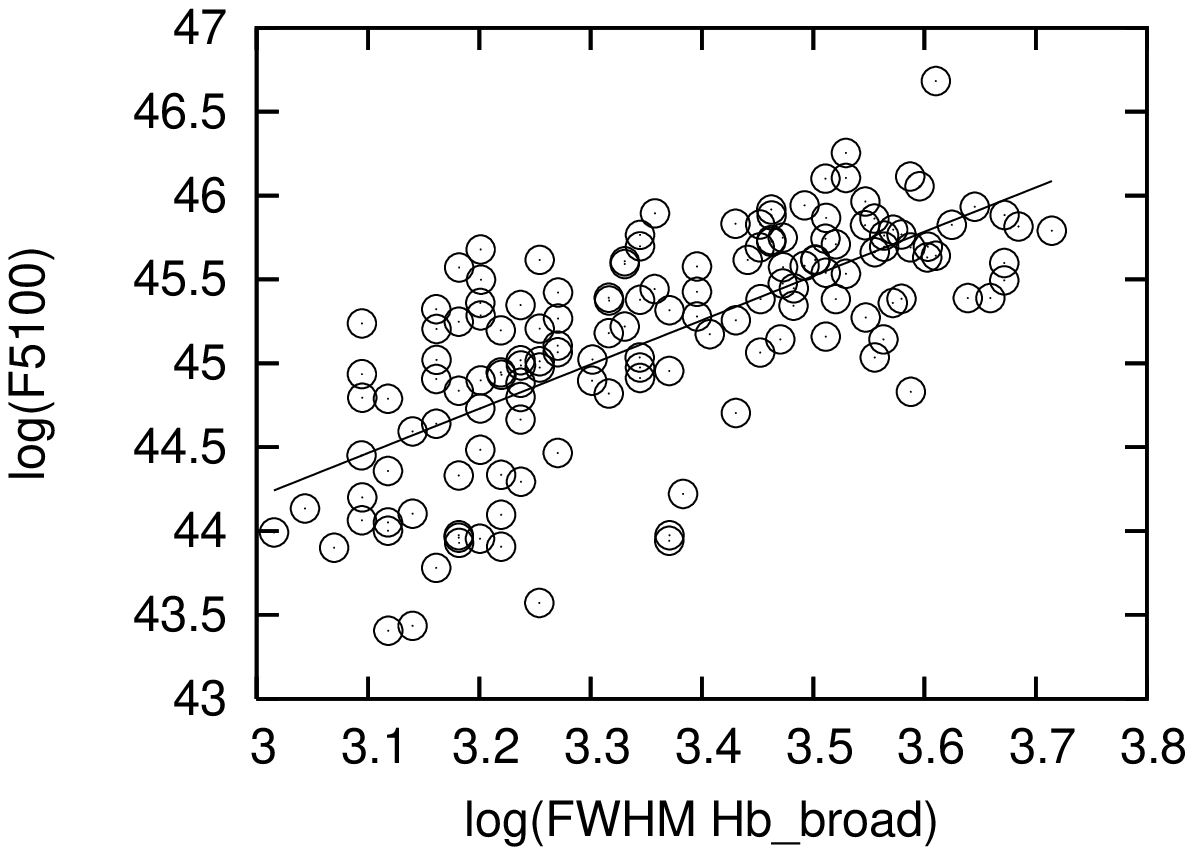}
\includegraphics[width=0.37\textwidth,angle=-90]{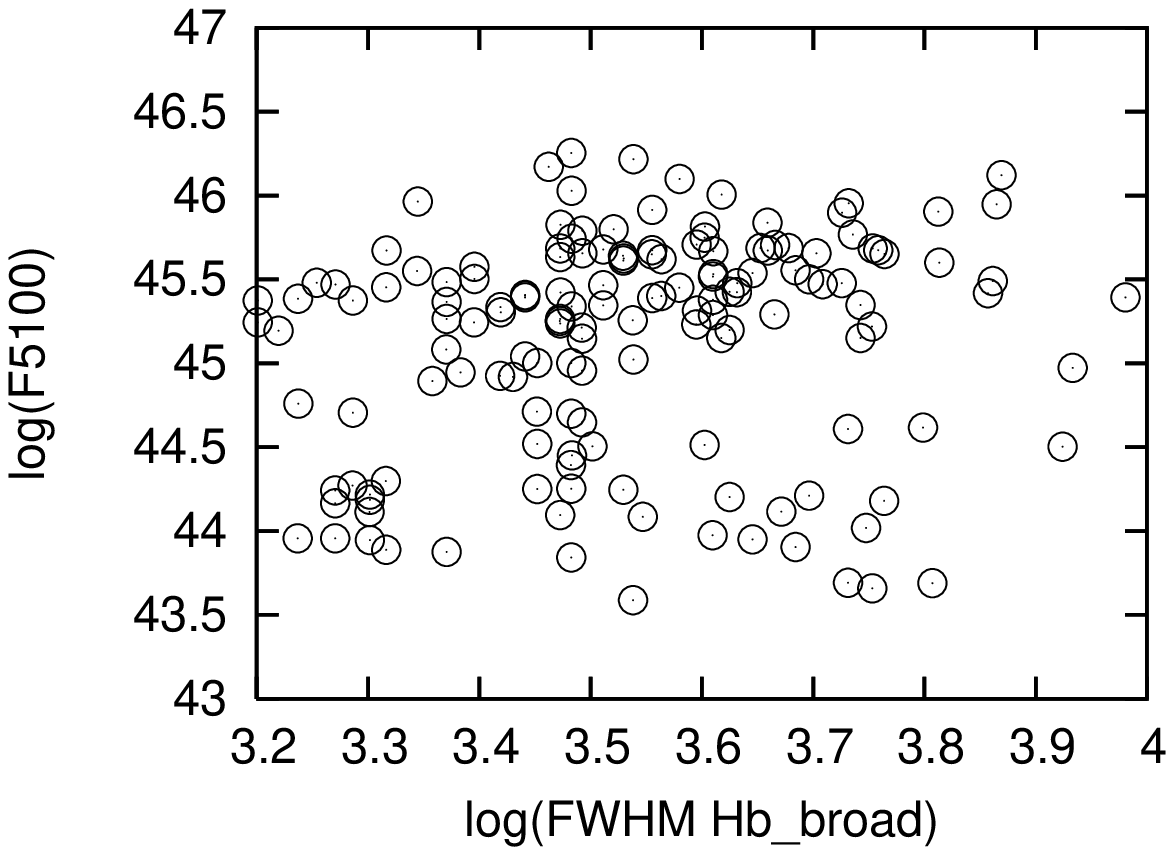}
\caption{Continuum luminosity against FWHM of the [\ion{O}{3}] and broad H$\beta$ lines for $R_1>-0.8$ (left) and $R_1<-0.8$ (right).}
\label{f67}
\end{figure}

\begin{figure}
\includegraphics[width=\textwidth]{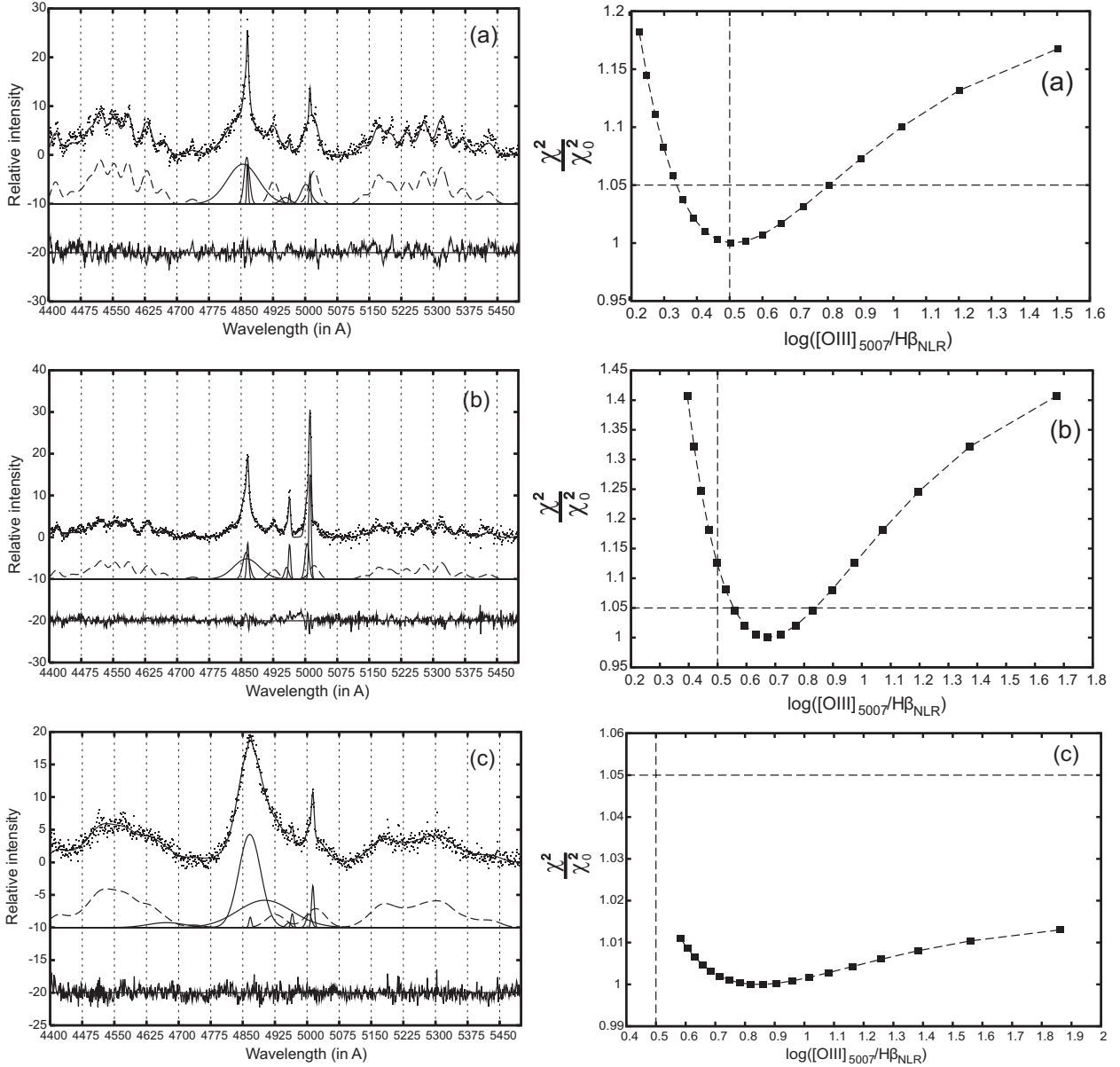}
\caption{Three characteristic spectra (left)  and corresponding $\chi^2/\chi_0^2$ (right) against $R$. The first panel shows the spectra where narrow H$\beta$ can be clearly seen.  The second shows where it is present but covered by the ILR component; and the third shows the case where the narrow line cannot be clearly seen. The horizontal line denotes $\chi^2/\chi_0^2$=1.05 and vertical line $R$=0.5.}
\label{fnew}
\end{figure}

\begin{figure}
\includegraphics[width=\textwidth]{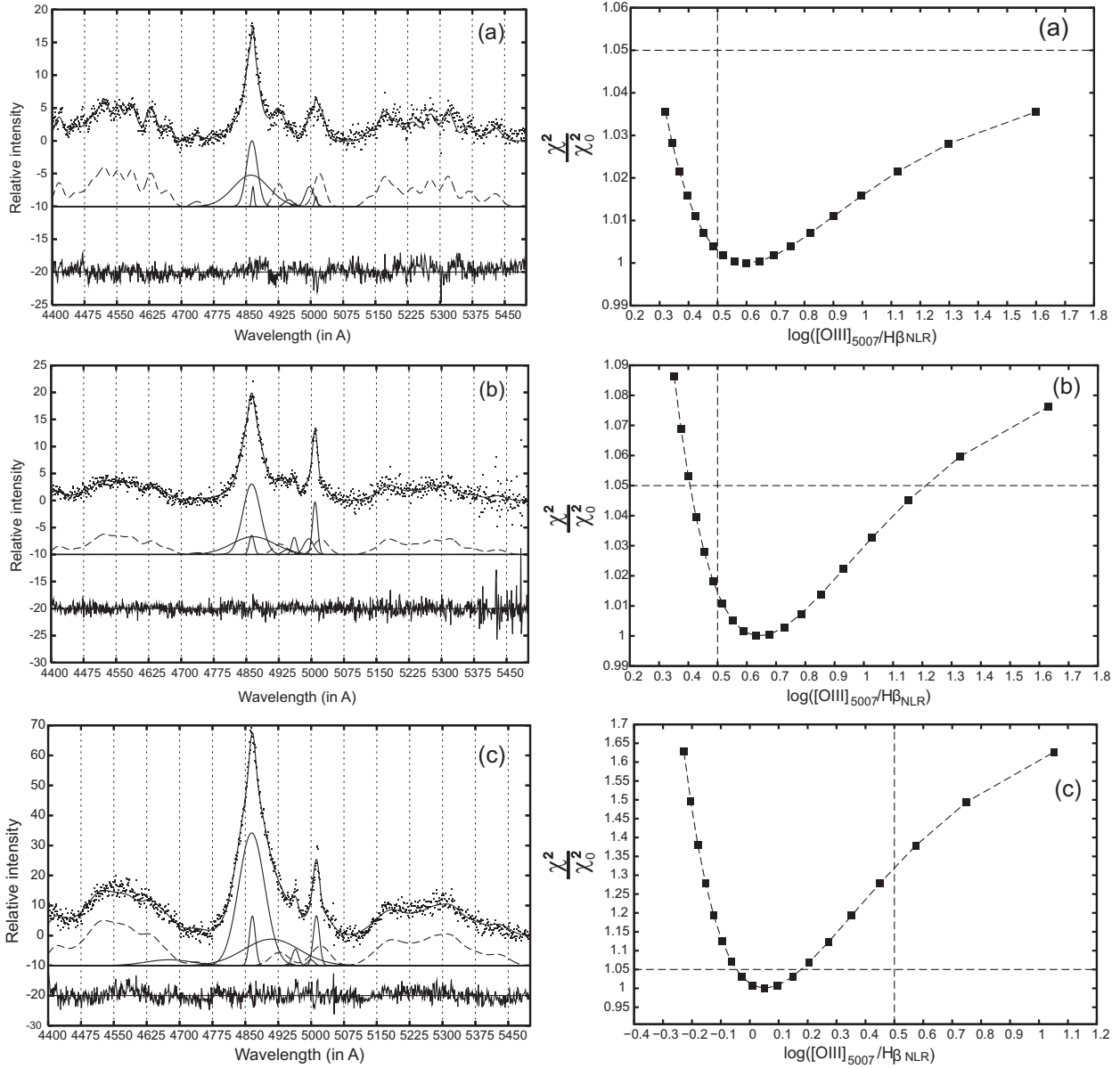}
\caption{The same as in Fig. \ref{fnew}, but for objects where the narrow H$\beta$ is not prominent: (a)  objects from the group with flat minima of $\chi^2$ ($\chi^2$ changes are less then 5\% for different values of the H$\beta$ NLR intensity), (b)  objects with $\chi^2$ changes higher than 5\%, but less then 10\%, and  (c) for objects with the very deep minimum of $\chi^2$ for  different values of the H$\beta$ NLR intensity (more than 10\%);}
\label{fnew1}
\end{figure}

\clearpage

\footnotesize
\begin{sidewaystable}[h]

\caption{Spearman rank order correlations for  the total sample and for subsamples: (1) log([O III]5007/H$\beta$ NLR)$>$ 0.5 and (2)log([O III]5007/H$\beta$ NLR)$<$ 0.5 .}
\begin{center}
\begin{tabular}{|c|cc|cc|cc|cc|cc|cc|cc|}

\tableline\tableline
&\multicolumn{2}{|c|}{{\tiny log($\lambda L_{5100}$)}}&\multicolumn{2}{|c|}{{\tiny   log(FWHM H$\beta$)}} & \multicolumn{2}{|c|}{{\tiny  log(EW [O III])}} &\multicolumn{2}{|c|} {{\tiny log(EW Fe II)}} &\multicolumn{2}{|c|}{{\tiny   log(EW H$\beta$ NLR)}} &\multicolumn{2}{|c|}{{\tiny  log(EW H$\beta$ broad)}} & \multicolumn{2}{|c|}{{\tiny log(FWMI10\%H$\beta$) }} \\
 &{\scriptsize r} &{\scriptsize  P }& {\scriptsize r }&{\scriptsize P} &{\scriptsize r }&{\scriptsize P }&{\scriptsize r }&{\scriptsize P }&{\scriptsize r }&{\scriptsize P }&{\scriptsize  r} &{\scriptsize P }&{\scriptsize r }&{\scriptsize P}\\
\tableline

 &\multicolumn{2}{|c|}{{\scriptsize    Total sample}  }&{\scriptsize  0.42 }&{\scriptsize  2.3E-14 }&{\scriptsize  -0.45 }&{\scriptsize  2.2E-16 }&{\scriptsize   0.27 }&{\scriptsize  1.6E-6 }&{\scriptsize  -0.41 }&{\scriptsize  9.4E-14 }&{\scriptsize  0.14 }&{\scriptsize  0.02 }&{\scriptsize  0.43 }&{\scriptsize  5.1E-15}\\
{\tiny log ($\lambda L_{5100}$) }& \multicolumn{2}{|c|}{{\tiny   (1)} }&{\scriptsize  0.26 }&{\scriptsize  1.1E-4 }&{\scriptsize  -0.51 }&{\scriptsize  3.6E-15 }&{\scriptsize  0.29 }&{\scriptsize  2.4E-5 }&{\scriptsize  -0.56 }&{\scriptsize  0 }&{\scriptsize  -0.03 }&{\scriptsize  0.67 }&{\scriptsize  0.29 }&{\scriptsize  2.3E-5}\\
 & \multicolumn{2}{|c|}{{\tiny   (2)} }&{\scriptsize  0.81 }&{\scriptsize  0 }&{\scriptsize  -0.46 }&{\scriptsize  4.2E-6 }&{\scriptsize  0.26 }&{\scriptsize  0.01 }&{\scriptsize  -0.27 }&{\scriptsize  0.01 }&{\scriptsize  0.56 }&{\scriptsize  9E-9 }&{\scriptsize  0.81 }&{\scriptsize  0}\\
 \tableline
 &{\scriptsize  0.42 }&{\scriptsize  2.3E-14 }& \multicolumn{2}{|c|}{{\scriptsize   total sample} }&{\scriptsize  -0.07 }&{\scriptsize  0.24 }&{\scriptsize  -0.24 }&{\scriptsize  2.3E-5  }&{\scriptsize  -0.34 }&{\scriptsize  1.5E-9

}&{\scriptsize  0.48 }&{\scriptsize  0 }&{\scriptsize  0.90 }&{\scriptsize  0}\\
 {\tiny   log(FWHM H$\beta$) }&{\scriptsize  0.26 }&{\scriptsize  1.1E-4 }&  \multicolumn{2}{|c|}{{\tiny   (1)} }&{\scriptsize  -0.08 }&{\scriptsize  0.24 }&{\scriptsize  -0.37 }&{\scriptsize  3.2E-8 }&{\scriptsize  -0.25 }&{\scriptsize  1.9E-4 }&{\scriptsize  0.38 }&{\scriptsize  1.2E-8 }&{\scriptsize  0.88 }&{\scriptsize  0}\\
  &{\scriptsize  0.81 }&{\scriptsize 0 }& \multicolumn{2}{|c|}{{\tiny   (2)} }&{\scriptsize  -0.42 }&{\scriptsize  4.1E-5 }&{\scriptsize  0.26 }&{\scriptsize  0.01 }&{\scriptsize  -0.15 }&{\scriptsize  0.16 }&{\scriptsize  0.53 }&{\scriptsize  4.8E-8 }&{\scriptsize  0.89 }&{\scriptsize  0}\\
 \tableline
 &{\scriptsize  -0.45 }&{\scriptsize  2.2E-16 }&{\scriptsize  -0.07 }&{\scriptsize  0.24 }& \multicolumn{2}{|c|}{{\scriptsize    total sample} }&{\scriptsize  -0.41 }&{\scriptsize  6.8E-14  }&{\scriptsize  0.32 }&{\scriptsize 2.1E-8 }&{\scriptsize  0.24 }&{\scriptsize  2.4E-5 }&{\scriptsize  -0.05 }&{\scriptsize  0.38}\\
  {\tiny log(EW [O III]) }&{\scriptsize  -0.51 }&{\scriptsize  3.6E-15 }&{\scriptsize  -0.08 }&{\scriptsize  0.24 }&\multicolumn{2}{|c|}{{\tiny   (1)} }&{\scriptsize  -0.38 }&{\scriptsize  7.9E-9 }&{\scriptsize  0.73 }&{\scriptsize  0 }&{\scriptsize  0.26 }&{\scriptsize 1.1E-4 }&{\scriptsize  -0.05}&{\scriptsize  0.51}\\
  &{\scriptsize  -0.46 }&{\scriptsize  4.2E-6 }&{\scriptsize  -0.42 }&{\scriptsize  4.1E-5 }& \multicolumn{2}{|c|}{{\tiny   (2)} }&{\scriptsize  -0.27 }&{\scriptsize  0.01 }&{\scriptsize  0.53 }&{\scriptsize  5.9E-8 }&{\scriptsize  -0.11 }&{\scriptsize  0.29 }&{\scriptsize  -0.48}&{\scriptsize  1.5E-6}\\
 \tableline
   &{\scriptsize  0.27 }&{\scriptsize  1.6E-6 }&{\scriptsize  -0.24 }&{\scriptsize  2.3E-5 }&{\scriptsize  -0.41 }&{\scriptsize  6.8E-14 }& \multicolumn{2}{|c|}{{\scriptsize    total sample} }&{\scriptsize  -0.01 }&{\scriptsize  0.80 }&{\scriptsize  0.07 }&{\scriptsize  0.23 }&{\scriptsize  -0.26}&{\scriptsize  6.3E-6}\\
 {\tiny log(EW Fe II)   }&{\scriptsize  0.29 }&{\scriptsize  2.4E-5 }&{\scriptsize  -0.37 }&{\scriptsize  3.2E-8 }&{\scriptsize  -0.38 }&{\scriptsize  7.9E-9 }& \multicolumn{2}{|c|}{{\tiny  (1)} }&{\scriptsize  -0.28}&{\scriptsize  3.9E-5 }&{\scriptsize  0.04 }&{\scriptsize  0.57 }&{\scriptsize  -0.38}&{\scriptsize  1.4E-8}\\
  &{\scriptsize  0.26 }&{\scriptsize  0.01 }&{\scriptsize  0.26 }&{\scriptsize  0.01 }&{\scriptsize  -0.27 }&{\scriptsize  0.01 }& \multicolumn{2}{|c|}{{\tiny   (2)} }&{\scriptsize  -0.04 }&{\scriptsize  0.67  }&{\scriptsize  0.49 }&{\scriptsize  9.4E-7 }&{\scriptsize  0.27}&{\scriptsize  0.01}\\
 \tableline
  &{\scriptsize  -0.41 }&{\scriptsize  9.4E-14 }&{\scriptsize  -0.34 }&{\scriptsize  1.5E-9 }&{\scriptsize  0.32 }&{\scriptsize  2.1E-8 }&{\scriptsize  -0.01 }&{\scriptsize  0.80 }& \multicolumn{2}{|c|}{{\scriptsize    total sample} }&{\scriptsize  -0.16 }&{\scriptsize  0.005 }&{\scriptsize  -0.36}&{\scriptsize  1.6E-10}\\
   {\tiny log(EW H$\beta$ NLR) }&{\scriptsize  -0.56 }&{\scriptsize  0 }&{\scriptsize  -0.25 }&{\scriptsize 1.9E-4 }&{\scriptsize  0.73 }&{\scriptsize  0 }&{\scriptsize  -0.28 }&{\scriptsize 3.9E-5 }& \multicolumn{2}{|c|}{{\tiny   (1)} }&{\scriptsize  0.08 }&{\scriptsize  0.23 }&{\scriptsize  -0.25 }&{\scriptsize  2.4E-4}\\
  &{\scriptsize  -0.27 }&{\scriptsize  0.01 }&{\scriptsize  -0.15 }&{\scriptsize  0.16 }&{\scriptsize  0.53 }&{\scriptsize  5.9E-8 }&{\scriptsize  -0.04 }&{\scriptsize  0.67 }& \multicolumn{2}{|c|}{{\tiny   (2)} }&{\scriptsize  -0.21 }&{\scriptsize  0.05 }&{\scriptsize  -0.21 }&{\scriptsize 0.05}  \\
 \tableline
  &{\scriptsize  0.14 }&{\scriptsize  0.02 }&{\scriptsize  0.48 }&{\scriptsize  0 }&{\scriptsize  0.24}&{\scriptsize  2.4E-5 }&{\scriptsize  0.07 }&{\scriptsize  0.23 }&{\scriptsize  -0.16 }&{\scriptsize  0.005 }& \multicolumn{2}{|c|}{{\scriptsize    total sample} }&{\scriptsize  0.50 }&{\scriptsize  0}\\
   {\tiny  log(EW H$\beta$ broad) }&{\scriptsize  -0.03 }&{\scriptsize  0.67 }&{\scriptsize  0.38 }&{\scriptsize  1.2E-8 }&{\scriptsize  0.26 }&{\scriptsize  1.1E-4 }&{\scriptsize  0.04 }&{\scriptsize  0.57 }&{\scriptsize  0.08 }&{\scriptsize  0.23 }& \multicolumn{2}{|c|}{{\tiny   (1)} }&{\scriptsize  0.42}&{\scriptsize  1.5E-10}\\
   &{\scriptsize  0.56 }&{\scriptsize  9E-9 }&{\scriptsize  0.53 }&{\scriptsize  4.8E-8 }&{\scriptsize  -0.11 }&{\scriptsize  0.29 }&{\scriptsize  0.49 }&{\scriptsize  9.4E-7 }&{\scriptsize  -0.21 }&{\scriptsize  0.05 }&  \multicolumn{2}{|c|}{{\tiny   (2)} }&{\scriptsize  0.49 }&{\scriptsize 6.4E-7}\\
 \tableline
 &{\scriptsize  0.43 }&{\scriptsize  5.1E-15 }&{\scriptsize  0.90}&{\scriptsize 0 }&{\scriptsize  -0.05 }&{\scriptsize  0.38 }&{\scriptsize  -0.26 }&{\scriptsize  6.3E-6}&{\scriptsize  -0.36 }&{\scriptsize  1.6E-10 }&{\scriptsize  0.50}&{\scriptsize  0 }& \multicolumn{2}{|c|}{{\scriptsize   total sample}}\\
{\tiny log(FWMI10\%H$\beta$) }&{\scriptsize  0.29}&{\scriptsize  2.3E-5}&{\scriptsize  0.88 }&{\scriptsize 0 }&{\scriptsize  -0.05 }&{\scriptsize  0.51 }&{\scriptsize  -0.38 }&{\scriptsize  1.4E-8 }&{\scriptsize  -0.25}&{\scriptsize  2.4E-4}&{\scriptsize  0.42 }&{\scriptsize 1.5E-10 }& \multicolumn{2}{|c|}{{\tiny   (1)}}\\
  &{\scriptsize  0.81 }&{\scriptsize  0 }&{\scriptsize  0.89 }&{\scriptsize  0 }&{\scriptsize  -0.48 }&{\scriptsize  1.5E-6 }&{\scriptsize  0.27 }&{\scriptsize  0.01 }&{\scriptsize  -0.21 }&{\scriptsize 0.05  }&{ \scriptsize 0.49 }&{\scriptsize  6.4E-7 }& \multicolumn{2}{|c|}{{\tiny   (2)}}\\
  \tableline
\end{tabular}
\end{center}
\label{t01}
\end{sidewaystable}

\clearpage

\clearpage

\footnotesize
\begin{sidewaystable}[h]

\caption{Spearman rank order correlations for the total sample and for subsamples: (1) log(FWHM [O III]narrow/FWHM H$\beta$)$<$ -0.8 and (2)log(FWHM [O III]/FWHM H$\beta$)$>$ -0.8 .}
\begin{center}
\begin{tabular}{|c|cc|cc|cc|cc|cc|cc|cc|}

\tableline\tableline
&\multicolumn{2}{|c|}{{\tiny log($\lambda L_{5100}$)}}&\multicolumn{2}{|c|}{{\tiny   log(FWHM H$\beta$)}} & \multicolumn{2}{|c|}{{\tiny  log(EW [O III])}} &\multicolumn{2}{|c|} {{\tiny log(EW Fe II)}} &\multicolumn{2}{|c|}{{\tiny   log(EW H$\beta$ NLR)}} &\multicolumn{2}{|c|}{{\tiny  log(EW H$\beta$ broad)}} & \multicolumn{2}{|c|}{{\tiny  log(FWMI10\%H$\beta$)}} \\
 &{\scriptsize r} &{\scriptsize  P }& {\scriptsize r }&{\scriptsize P} &{\scriptsize r }&{\scriptsize P }&{\scriptsize r }&{\scriptsize P }&{\scriptsize r }&{\scriptsize P }&{\scriptsize  r} &{\scriptsize P }&{\scriptsize r }&{\scriptsize P}\\
\tableline

 &\multicolumn{2}{|c|}{{\scriptsize    Total sample}  }&{\scriptsize  0.42 }&{\scriptsize  2.3E-14 }&{\scriptsize  -0.45 }&{\scriptsize  2.2E-16 }&{\scriptsize   0.27 }&{\scriptsize  1.6E-6 }&{\scriptsize  -0.41 }&{\scriptsize  9.4E-14 }&{\scriptsize  0.14 }&{\scriptsize  0.02 }&{\scriptsize  0.43 }&{\scriptsize  5.1E-15}\\
{\tiny log ($\lambda L_{5100}$) }& \multicolumn{2}{|c|}{{\tiny   (1)} }&{\scriptsize  0.21 }&{\scriptsize  0.01 }&{\scriptsize  -0.45 }&{\scriptsize  2.1E-8 }&{\scriptsize  0.25 }&{\scriptsize  0.003 }&{\scriptsize  -0.56 }&{\scriptsize  1.2E-12 }&{\scriptsize  -0.08 }&{\scriptsize  0.35 }&{\scriptsize  0.23 }&{\scriptsize  0.005}\\
 & \multicolumn{2}{|c|}{{\tiny   (2)} }&{\scriptsize  0.74 }&{\scriptsize  0 }&{\scriptsize  -0.44 }&{\scriptsize  2.5E-8 }&{\scriptsize  0.32 }&{\scriptsize  8.0E-5 }&{\scriptsize  -0.37 }&{\scriptsize  5.9E-6 }&{\scriptsize  0.34 }&{\scriptsize  3.1E-5 }&{\scriptsize  0.74 }&{\scriptsize  0}\\
 \tableline
 &{\scriptsize  0.42 }&{\scriptsize  2.3E-14 }& \multicolumn{2}{|c|}{{\scriptsize   total sample} }&{\scriptsize  -0.07 }&{\scriptsize  0.24 }&{\scriptsize  -0.24 }&{\scriptsize  2.3E-5  }&{\scriptsize  -0.34 }&{\scriptsize  1.5E-9
 }&{\scriptsize  0.48 }&{\scriptsize  0 }&{\scriptsize  0.90 }&{\scriptsize  0}\\
 {\tiny   log(FWHM H$\beta$) }&{\scriptsize  0.21 }&{\scriptsize  0.01 }&  \multicolumn{2}{|c|}{{\tiny   (1)} }&{\scriptsize  0.09 }&{\scriptsize  0.29 }&{\scriptsize  -0.41 }&{\scriptsize  5.9E-7 }&{\scriptsize  -0.12 }&{\scriptsize  0.14}&{\scriptsize  0.32 }&{\scriptsize  1.1E-4 }&{\scriptsize  0.87 }&{\scriptsize  0}\\
  &{\scriptsize  0.74 }&{\scriptsize 0 }& \multicolumn{2}{|c|}{{\tiny   (2)} }&{\scriptsize  -0.32 }&{\scriptsize  1.0E-4 }&{\scriptsize  0.24 }&{\scriptsize  0.003 }&{\scriptsize  -0.21 }&{\scriptsize  0.009}&{\scriptsize  0.50 }&{\scriptsize  1.6E-10 }&{\scriptsize  0.87 }&{\scriptsize  0}\\
 \tableline
 &{\scriptsize  -0.45 }&{\scriptsize  2.2E-16 }&{\scriptsize  -0.07 }&{\scriptsize  0.24 }& \multicolumn{2}{|c|}{{\scriptsize    total sample} }&{\scriptsize  -0.41 }&{\scriptsize  6.9E-14  }&{\scriptsize  0.32 }&{\scriptsize 2.1E-8 }&{\scriptsize  0.24 }&{\scriptsize  2.4E-5 }&{\scriptsize  -0.05 }&{\scriptsize  0.38}\\
  {\tiny log(EW [O III]) }&{\scriptsize  -0.45 }&{\scriptsize  2.1E-8 }&{\scriptsize  0.09 }&{\scriptsize  0.29 }&\multicolumn{2}{|c|}{{\tiny   (1)} }&{\scriptsize  -0.42 }&{\scriptsize  3.0E-7 }&{\scriptsize  0.71 }&{\scriptsize  0 }&{\scriptsize  0.46 }&{\scriptsize 1.7E-8 }&{\scriptsize  0.13}&{\scriptsize  0.13}\\
  &{\scriptsize  -0.44 }&{\scriptsize  2.5E-8 }&{\scriptsize  -0.32 }&{\scriptsize  1.0E-4 }& \multicolumn{2}{|c|}{{\tiny   (2)} }&{\scriptsize  -0.43 }&{\scriptsize  7.2E-8 }&{\scriptsize  0.18 }&{\scriptsize  0.03 }&{\scriptsize  0.02 }&{\scriptsize  0.81 }&{\scriptsize  -0.33}&{\scriptsize  6.4E-5}\\
 \tableline
   &{\scriptsize  0.27 }&{\scriptsize  1.6E-6 }&{\scriptsize  -0.24 }&{\scriptsize  2.3E-5 }&{\scriptsize  -0.41 }&{\scriptsize  6.9E-14 }& \multicolumn{2}{|c|}{{\scriptsize    total sample} }&{\scriptsize  -0.01 }&{\scriptsize  0.80 }&{\scriptsize  0.07 }&{\scriptsize  0.23 }&{\scriptsize  -0.26}&{\scriptsize  6.3E-6}\\
 {\tiny log(EW Fe II)   }&{\scriptsize  0.25 }&{\scriptsize  0.003 }&{\scriptsize  -0.41 }&{\scriptsize  5.9E-7 }&{\scriptsize  -0.42 }&{\scriptsize  3.0E-7 }& \multicolumn{2}{|c|}{{\tiny  (1)} }&{\scriptsize  -0.36}&{\scriptsize  1.1E-5 }&{\scriptsize  0.05 }&{\scriptsize  0.55 }&{\scriptsize  -0.42}&{\scriptsize  2.0E-7}\\
  &{\scriptsize  0.32 }&{\scriptsize  8.0E-5 }&{\scriptsize  0.24 }&{\scriptsize  0.003 }&{\scriptsize  -0.43 }&{\scriptsize  7.2E-8 }& \multicolumn{2}{|c|}{{\tiny   (2)} }&{\scriptsize  -0.04 }&{\scriptsize  0.65  }&{\scriptsize  0.36 }&{\scriptsize  6.7E-6 }&{\scriptsize  0.18}&{\scriptsize  0.03}\\
 \tableline
  &{\scriptsize  -0.41 }&{\scriptsize  9.4E-14 }&{\scriptsize  -0.34 }&{\scriptsize  1.5E-9 }&{\scriptsize  0.32 }&{\scriptsize  2.1E-8 }&{\scriptsize  -0.01 }&{\scriptsize  0.80 }& \multicolumn{2}{|c|}{{\scriptsize    total sample} }&{\scriptsize  -0.16 }&{\scriptsize  0.005 }&{\scriptsize  -0.36}&{\scriptsize  1.6E-10}\\
   {\tiny log(EW H$\beta$ NLR) }&{\scriptsize  -0.56 }&{\scriptsize  1.2E-12 }&{\scriptsize  -0.12 }&{\scriptsize 0.14 }&{\scriptsize  0.71 }&{\scriptsize  0 }&{\scriptsize  -0.36 }&{\scriptsize 1.1E-5 }& \multicolumn{2}{|c|}{{\tiny   (1)} }&{\scriptsize  0.15 }&{\scriptsize  0.08 }&{\scriptsize  -0.15 }&{\scriptsize  0.08}\\
  &{\scriptsize  -0.37 }&{\scriptsize  5.9E-6 }&{\scriptsize  -0.21 }&{\scriptsize  0.009 }&{\scriptsize  0.18 }&{\scriptsize  0.03 }&{\scriptsize  -0.04 }&{\scriptsize  0.65 }& \multicolumn{2}{|c|}{{\tiny   (2)} }&{\scriptsize  -0.19 }&{\scriptsize  0.02 }&{\scriptsize  -0.28 }&{\scriptsize 6.8E-4}  \\
 \tableline
  &{\scriptsize  0.14 }&{\scriptsize  0.02 }&{\scriptsize  0.48 }&{\scriptsize  0 }&{\scriptsize  0.24}&{\scriptsize  2.4E-5 }&{\scriptsize  0.07 }&{\scriptsize  0.23 }&{\scriptsize  -0.16 }&{\scriptsize  0.005 }& \multicolumn{2}{|c|}{{\scriptsize    total sample} }&{\scriptsize  0.50 }&{\scriptsize  0}\\
   {\tiny  log(EW H$\beta$ broad) }&{\scriptsize  -0.08 }&{\scriptsize  0.35 }&{\scriptsize  0.32 }&{\scriptsize  1.1E-4 }&{\scriptsize  0.46 }&{\scriptsize  1.7E-8 }&{\scriptsize  0.05 }&{\scriptsize  0.55 }&{\scriptsize  0.15 }&{\scriptsize  0.08 }& \multicolumn{2}{|c|}{{\tiny   (1)} }&{\scriptsize  0.38 }&{\scriptsize  4.3E-6}\\
   &{\scriptsize  0.34 }&{\scriptsize  3.1E-5 }&{\scriptsize  0.50 }&{\scriptsize  1.6E-10 }&{\scriptsize  0.02 }&{\scriptsize  0.81 }&{\scriptsize  0.36 }&{\scriptsize  6.7E-6 }&{\scriptsize  -0.19 }&{\scriptsize  0.02 }&  \multicolumn{2}{|c|}{{\tiny   (2)} }&{\scriptsize  0.46 }&{\scriptsize  4.9E-9}\\
 \tableline
 &{\scriptsize  0.43 }&{\scriptsize  5.1E-15 }&{\scriptsize  0.90 }&{\scriptsize  0 }&{\scriptsize  -0.05 }&{\scriptsize  0.38 }&{\scriptsize  -0.26 }&{\scriptsize  6.3E-6}&{\scriptsize  -0.36 }&{\scriptsize  1.6E-10 }&{\scriptsize  0.50}&{\scriptsize  0 }& \multicolumn{2}{|c|}{{\scriptsize   total sample}}\\
{\tiny log(FWMI10\%H$\beta$)  }&{\scriptsize  0.23}&{\scriptsize  0.005}&{\scriptsize  0.87 }&{\scriptsize 0 }&{\scriptsize  0.13 }&{\scriptsize  0.13 }&{\scriptsize  -0.42 }&{\scriptsize  2.0E-7 }&{\scriptsize  -0.15 }&{\scriptsize  0.08}&{\scriptsize  0.38 }&{\scriptsize 4.3E-6 }& \multicolumn{2}{|c|}{{\tiny   (1)}}\\
  &{\scriptsize  0.74 }&{\scriptsize  0 }&{\scriptsize  0.87 }&{\scriptsize  0 }&{\scriptsize  -0.33 }&{\scriptsize  6.4E-5 }&{\scriptsize  0.18 }&{\scriptsize  0.03 }&{\scriptsize  -0.28 }&{\scriptsize 6.8E-4  }&{\scriptsize 0.46 }&{\scriptsize  4.9E-9 }& \multicolumn{2}{|c|}{{\tiny   (2)}}\\
  \tableline
\end{tabular}
\end{center}
\label{t02}
\end{sidewaystable}

\begin{table}
\begin{center}
\caption{Principal component analysis of subsamples dependent of [\ion{O}{3}]/H$\beta_{NLR}$  and total sample. \label{t03}}

\begin{tabular}{|c|c c c|}

\tableline\tableline
\footnotesize{log([\ion{O}{3}]/H$\beta_{NLR}>$0.5 subsample}&Comp.1&Comp.2&Comp.3    \\
\tableline
\vspace*{-0.1cm} Standard deviation   & 1.519&1.237 &1.046 \\
\vspace*{-0.1cm}Proportion of Variance &0.385 &0.255& 0.182\\
\vspace*{-0.1cm}Cumulative Proportion  &0.385& 0.640& 0.822\\
 \tableline
\vspace*{-0.1cm}$\lambda L_{5100}$ &-0.523& 0.088 &0.009\\
\vspace*{-0.1cm}FWHM H$\beta$      &-0.154& 0.737 & -0.135\\
\vspace*{-0.1cm}EW [\ion{O}{3}]         & 0.571& 0.138 &0.119\\
\vspace*{-0.1cm}EW \ion{Fe}{2}           &-0.307&-0.382 &0.676\\
\vspace*{-0.1cm}EW H$\beta$ NLR    & 0.529&-0.124 &0.158\\
\vspace*{-0.1cm}EW H$\beta$ broad    & 0.057& 0.518 &0.696\\
  \tableline\tableline

\footnotesize{log([\ion{O}{3}]/H$\beta_{NLR}<$0.5 subsample}&Comp.1&Comp.2&Comp.3  \\
\tableline
\vspace*{-0.1cm} Standard deviation    &1.706&1.157&0.903\\
\vspace*{-0.1cm}Proportion of Variance &0.485&0.223&0.136 \\
\vspace*{-0.1cm}Cumulative Proportion  &0.485&0.708&0.844 \\
 \tableline
\vspace*{-0.1cm}$\lambda L_{5100}$ &-0.512&-0.044& -0.377\\
\vspace*{-0.1cm}FWHM H$\beta$      &-0.490&-0.111& -0.446\\
\vspace*{-0.1cm}EW [\ion{O}{3}]         & 0.384&-0.532& -0.160\\
\vspace*{-0.1cm}EW \ion{Fe}{2}           &-0.331&-0.326&0.758\\
\vspace*{-0.1cm}EW H$\beta$ NLR    & 0.277&-0.613& -0.225\\
\vspace*{-0.1cm}EW H$\beta$ broad    &-0.404&-0.469& 0.091\\

  \tableline\tableline

\footnotesize{total sample}&Comp.1&Comp.2&Comp.3   \\
\tableline
 \vspace*{-0.1cm}Standard deviation   &  1.458&1.272 &1.023 \\
\vspace*{-0.1cm}Proportion of Variance &0.354 &0.270& 0.174\\
\vspace*{-0.1cm}Cumulative Proportion  &0.354& 0.624& 0.798\\
 \tableline
\vspace*{-0.1cm}$\lambda L_{5100}$ &-0.570& -0.067& 0.004\\
\vspace*{-0.1cm}FWHM H$\beta$&-0.394&  0.544& -0.117\\
\vspace*{-0.1cm}EW [\ion{O}{3}]& 0.454&  0.458&  0.178\\
\vspace*{-0.1cm}EW \ion{Fe}{2}   &-0.246& -0.471&  0.651\\
\vspace*{-0.1cm}EW H$\beta$ NLR & 0.459& -0.090&  0.363\\
\vspace*{-0.1cm}EW H$\beta$ broad &-0.208&  0.510&  0.631\\
  \tableline

\end{tabular}
\end{center}
\end{table}

  \begin{table}
\begin{center}
\caption{Principal component analysis of subsamples dependent of log(FWHM [\ion{O}{3}]/FWHM H$\beta$) ratio. \label{t04}}

\begin{tabular}{|c|c c c|}

\tableline\tableline
\footnotesize{log(FWHM [\ion{O}{3}]/FWHM H$\beta$)$<$ -0.8 subsample}&Comp.1&Comp.2&Comp.3    \\
\tableline
\vspace*{-0.1cm} Standard deviation   & 1.531&1.203 &1.070 \\
\vspace*{-0.1cm}Proportion of Variance &0.391 &0.241& 0.191\\
\vspace*{-0.1cm}Cumulative Proportion  &0.391& 0.632& 0.823\\
 \tableline
\vspace*{-0.1cm}$\lambda L_{5100}$ &-0.481& -0.248 &0.030\\
\vspace*{-0.1cm}FWHM H$\beta$      & 0.058& -0.763 & -0.194\\
\vspace*{-0.1cm}EW [\ion{O}{3}]         & 0.585& -0.067 &0.138\\
\vspace*{-0.1cm}EW \ion{Fe}{2}           &-0.362&0.225 &0.686\\
\vspace*{-0.1cm}EW H$\beta$ NLR    & 0.501&0.281 &0.116\\
\vspace*{-0.1cm}EW H$\beta$ broad    & 0.200& -0.471 &0.677\\
  \tableline\tableline

\footnotesize{log(FWHM [\ion{O}{3}]/FWHM H$\beta$)$>$ -0.8 subsample}&Comp.1&Comp.2&Comp.3  \\
\tableline
\vspace*{-0.1cm} Standard deviation    &1.577&1.047&1.004\\
\vspace*{-0.1cm}Proportion of Variance &0.414&0.183&0.168 \\
\vspace*{-0.1cm}Cumulative Proportion  &0.414&0.597&0.765 \\
 \tableline
\vspace*{-0.1cm}$\lambda L_{5100}$ &-0.530&-0.014& 0.245\\
\vspace*{-0.1cm}FWHM H$\beta$      &-0.530&0.137& 0.065\\
\vspace*{-0.1cm}EW [\ion{O}{3}]         & 0.302&0.763& -0.083\\
\vspace*{-0.1cm}EW \ion{Fe}{2}           &-0.365&-0.265&-0.514\\
\vspace*{-0.1cm}EW H$\beta$ NLR    & 0.266&-0.250& -0.664\\
\vspace*{-0.1cm}EW H$\beta$ broad    &-0.378&0.516& -0.472\\

  \tableline

\end{tabular}
\end{center}
\end{table}

   \end{document}